\documentclass[usenatbib]{mnras}

\usepackage{graphicx}
\usepackage{hyperref}
\usepackage{xspace}
\usepackage{amsmath}

\def\mobs{M_*^{\mathrm{(obs)}}}

\def\mstar{M_*}

\def\mchab{M_*^{\mathrm{(Chab)}}}

\def\mhalo{M_{h}}

\def\mbh{M_{\mathrm{BH}}}

\def\chalo{c_{h}}

\def\aimf{\alpha_{\mathrm{IMF}}}

\def\reff{R_{\mathrm{e}}}
\def\rein{R_{\mathrm{Ein}}}

\def\sigmaap{\sigma_{\mathrm{ap}}}

\def\hyperp{\boldsymbol{\eta}}
\def\individ{\boldsymbol{\psi}}
\def\individi{\boldsymbol{\psi}_i}

\def\data{\mathbf{d}}
\def\datai{\mathbf{d}_i}
\def\wl{\left\{\rm{WL}\right\}}

\def\Sref#1{Section~\ref{#1}\xspace}
\def\Fref#1{Figure~\ref{#1}\xspace}
\def\Tref#1{Table~\ref{#1}\xspace}
\def\Eref#1{Equation~\ref{#1}\xspace}

\def\pr{{\rm P}}

\begin{document}

\title[Radial variations in stellar mass-to-light ratio]{Evidence for radial variations in the stellar mass-to-light ratio of massive galaxies from weak and strong lensing}

\author[Sonnenfeld, Leauthaud, Auger, Gavazzi, Treu, More, Komiyama]{
Alessandro~Sonnenfeld,$^{1}$\thanks{E-mail:alessandro.sonnenfeld@ipmu.jp}
Alexie~Leauthaud$^{1, 2}$,
Matthew~W.~Auger$^{3}$,
\newauthor
Raphael~Gavazzi$^{4}$,
Tommaso~Treu$^{5}$,
Surhud~More$^{1}$,
Yutaka~Komiyama$^{6, 7}$
\\
$^{1}$Kavli IPMU (WPI), UTIAS, The University of Tokyo, Kashiwa, Chiba 277-8583, Japan \\
$^{2}$ Department of Astronomy and Astrophysics, UCO/Lick Observatory, University of California, 1156 High Street, Santa Cruz, CA 95064, USA\\
$^{3}$ Institute of Astronomy, Madingley Road, Cambridge CB3 0HA, UK \\
$^{4}$ Institut d’Astrophysique de Paris, UMR 7095 CNRS \& Université Pierre et Marie Curie, 98 bis Bd Arago, 75014 Paris, France \\
$^{5}$ Department of Physics and Astronomy, 430 Portola Plaza, Los Angeles, CA 90095-1547, USA \\
$^{6}$ National Astronomical Observatory of Japan, 2-21-1 Osawa, Mitaka, Tokyo 181-8588, Japan \\
$^{7}$ SOKENDAI(The Graduate University for Advanced Studies), Mitaka,
 Tokyo, 181-8588, Japan
}

\maketitle

\begin{abstract}
The Initial Mass Function (IMF) for massive galaxies can be constrained by combining stellar dynamics with strong gravitational lensing. However, this method is limited by degeneracies between the density profile of dark matter and the stellar mass-to-light ratio. In this work we reduce this degeneracy by combining weak lensing together with strong lensing and stellar kinematics. Our analysis is based on two galaxy samples: 45 strong lenses from the SLACS survey and 1,700 massive quiescent galaxies from the SDSS main spectroscopic sample with weak lensing measurements from the Hyper Suprime-Cam survey. We use a Bayesian hierarchical approach to jointly model all three observables. We fit the data with models of varying complexity and show that a model with a radial gradient in the stellar mass-to-light ratio is required to simultaneously describe both galaxy samples. 
This result is driven by a subset of strong lenses with very steep total density profile, that cannot be fitted by models with no gradient.
Our measurements are unable to determine whether $M_*/L$ gradients are due to variations in stellar population parameters at fixed IMF, or to gradients in the IMF itself. 
The inclusion of $M_*/L$ gradients decreases dramatically the inferred IMF normalisation, compared to previous lensing-based studies, with the exact value depending on the assumed dark matter profile. 
The main effect of strong lensing selection is to shift the stellar mass distribution towards the high mass end, while the halo mass and stellar IMF distribution at fixed stellar mass are not significantly affected.
\end{abstract}


\section{Introduction}\label{sect:intro}

The determination of the stellar initial mass function (IMF) of massive early-type galaxies (ETGs) is an outstanding problems in astrophysics. Better constraints on the stellar IMF of massive galaxies will help to improve our understanding of star formation in extreme environments, such as the cores of massive galaxies.
Moreover, the stellar IMF is currently the main systematic uncertainty in the determination of galaxy stellar masses. This limits our ability to match the numerical simulations, where the most precise predictions are made for {\em galaxy mass}, and the real Universe, where the most robust observations are those regarding {\em galaxy light}.
Similarly, our ability to compare theoretical predictions with observations is limited by our ignorance of the dark matter distribution inside and around galaxies.

In recent years there have been a large number of studies aimed at determining the stellar IMF of massive ETGs \citep[][and references therein]{Tre++10, v+C10, Cap++12, Fer++13, Dut++13, Spi++14}.
One of the observational tools that can be used to measure the IMF of galaxies and their dark matter content at cosmological distances is strong gravitational lensing.
In a fundamental study, \citet{Aug++10b} explored the degeneracy between the choice of the IMF and the parametrization of the density profile of dark matter halo using strong lensing, weak lensing, and stellar kinematics data for a sample of 59 galaxies from the Sloan Lens ACS Survey \citep[SLACS][]{Bol++06, Aug++10}. \citet{Aug++10b} found that an IMF normalisation slightly heavier than that of a Salpeter IMF and increasing with stellar mass are required to fit the data if a Navarro Frenk \& White \citep[NFW,][]{NFW97} profile is assumed.
In this work we revisit that conclusion. 

We use the same data as \citet{Aug++10b} work and complement it with weak lensing and stellar kinematics observations for a set of 1,700 massive galaxies from the Sloan Digital Sky Survey (SDSS) spectroscopic sample and shear data from the Hyper Suprime-Cam Subaru Strategic Program \citep[HSC SSP][hereafter `HSC survey']{Aih++18}.
Our goal is to find the simplest model for the distribution of dark matter and the IMF of massive galaxies that can simultaneously reproduce all of our observations.

Our weak lensing measurements greatly help to pin down the mean halo mass of the sample, thereby allowing us to partially break the degeneracy between dark matter density profile and stellar IMF. However, it also introduces an additional challenge. The sample of SLACS lenses is too small to obtain high signal-to-noise measurements of weak lensing. We therefore use a large sample of galaxies which are drawn from a similar population as the strong lensing sample, but which are themselves not necessarily strong lensing systems. In order to model both samples simultaneously, we need to model the selection effects of the strong lensing sample. 

We adopt the following strategy. We use a Bayesian hierarchical inference method to infer the distribution of halo masses and the normalisation of the IMF as a function of stellar mass for SLACS lenses and HSC galaxies independently. We then use the model that best describes the HSC data to generate a mock population of galaxies. We apply a strong lensing selection to this mock sample, and compare its properties with those inferred from the SLACS sample. If the assumed model is accurate, then the prediction from both the HSC sample, and the SLACS sample should match up. We explore three different models, with increasing degrees of complexity,
until we reach such an agreement.

This paper is organised as follows. In \Sref{sect:data} we present the data. In \Sref{sect:model} we introduce our Bayesian hierarchical inference method. In \Sref{sect:results} we fit three different models to the data. 
We discuss our results in \Sref{sect:discuss} and conclude in \Sref{sect:concl}.
We assume a flat $\Lambda$CDM cosmology with $\Omega_M=0.3$ and $H_0=70\,\rm{km}\,\rm{s}^{-1}\,\rm{Mpc}^{-1}$.
Halo masses, labelled as $\mhalo$, are defined as the mass enclosed within a spherical shell with average density equal to $200$ times the critical density of the Universe. Halo radius is labelled $R_h$ and is also defined assuming a boundary defined with respect to $200$ times the critical density. Halo concentration is labelled as $\chalo$ and is defined as $\chalo=R_s/R_h$ where $R_s$ is the halo scale radius. Stellar and halo masses are expressed in units of solar mass, while sizes and projected distances are expressed in physical units.

\section{The data}\label{sect:data}

\subsection{SLACS strong lensing}

Our sample of strong lens systems is drawn from the set of 59 SLACS lenses studied by \citet{Aug++10}.
The selection of SLACS lenses is described in detail by \citet{Bol++06}, and is briefly summarised here.
SLACS lenses were discovered by scanning spectra of quiescent galaxies from the SDSS main spectroscopic sample and looking for multiple emission lines associated with objects at a higher redshift than the main target galaxy. Strong lens candidates obtained with this method were followed-up with high resolution imaging from Hubble Space Telescope (HST) to confirm their lens nature and to measure their properties.

\citet{Aug++09} present a sample of 84 strong lenses discovered with this procedure. Of these lenses, 6 are morphologically classified as disk galaxies, and the remaining are classified as early-type galaxies. The \citet{Aug++10} sample, from which we select the lenses for our study, consists of a subset of 59 lenses drawn from the \citet{Aug++09} sample. These have been selected to have early-type morphology, robust lens models, and reliable stellar kinematics measurements.

For each of these 59 lenses, the following data is available:
\begin{itemize}
\item Lens and source redshifts and lens velocity dispersions within the SDSS spectroscopic fibre  $\sigmaap$ \citep[][]{Aug++09}.
\item Rest-frame I-band effective radii obtained from de Vaucouleurs profile fit to multi-band HST imaging data \citep[][]{Aug++09}.
\item Stellar masses obtained from spectral energy distribution (SED) fitting to HST imaging data, assuming a Chabrier IMF, $\mchab$ \citep{Aug++09}.
\item Einstein radii $\rein$ \citep{Aug++10}.
\end{itemize}
In addition, for a subset of 33 galaxies, weak lensing shear measurements, obtained with deep data taken with the Advanced Camera for Surveys on HST, are available. We refer to \citet{Gav++07} for a description of these weak lensing measurements.

\subsection{HSC massive galaxies}

The Hyper Suprime-Cam \citep[HSC,][]{Miy++17, Kom++17, Kaw++17, Fur++17} is an optical camera with a $1.5$~deg$^2$ field of view, installed on the Subaru Telescope.
The HSC survey \citep{Aih++18} is an ongoing 5 band $grizy$ photometric survey. The wide layer will cover $1,400$~deg$^2$ to $26.2$~mag and with a mean seeing of $\sim0.6''$ in the $i$-band.

We wish to build a sample of galaxies selected in a similar way as the parent sample used to build the SLACS sample, in the region of the sky covered by HSC data.
We start from the SDSS main spectroscopic sample and select objects with a quiescent spectrum by imposing a maximum H-$\alpha$ emission equivalent width of $3\AA$.
We then eliminate highly elongated objects by selecting objects with a ratio between minor and major axis larger than $0.5$, as measured by a fit of a de Vaucouleurs profile \citep{deV48} to the SDSS $r$-band.
This second step is designed to approximate the selection of objects with an early-type morphology, as done by \citet{Aug++10}. 
Since we plan to use weak lensing measurements for this sample of galaxies, we also apply a cut in redshift, selecting only objects at $z > 0.05$, so that each galaxy contributes with a non-negligible lensing signal.
For the resulting sample of galaxies, we then consider the following data, 
\begin{itemize}
\item Stellar masses from the MPA JHU catalogue \citep{Kau++03}, obtained through stellar population synthesis fits to SDSS broadband photometry and assuming a Chabrier IMF.
\item Half-light radii obtained from a de Vaucouleurs profile fit to the SDSS $r$-band as provided by SDSS DR12 \citep{Ala++15}.
\item Stellar velocity dispersion measurements within the SDSS spectroscopic fibre.
\item Weak lensing measurements from HSC (described in the next subsection).
\end{itemize}
Shape measurements and photo-z measurements for background weak lensing source galaxies are the only data from the HSC survey that we use in this work. Galaxy sizes and stellar masses are obtained from SDSS. Although HSC photometry is deeper and has better image quality than the SDSS, we do not expect it to bring a significant improvement on the measurements of our sample, since these low-redshift massive galaxies are well resolved and detected with high signal-to-noise in SDSS data.

We wish to carry out a weak lensing analysis around the galaxies in this sample, using HSC shear data. We will be using the Bayesian hierarchical inference method developed by \citet{S+L18}. The \citet{S+L18} method relies on a critical assumption, that of {\em isolated lenses}: lens galaxies are assumed to be at an infinite projected distance from each other, so that each background source is only lensed by one lens, at most.
This approximation was shown by \citet{S+L18} to be valid in the high mass regime, around lens galaxies with $\log{\mstar} > 11$, but likely breaks down at low masses. We therefore apply a cut in stellar mass to our sample and select galaxies with observed stellar mass larger than $10^{11}M_\odot$. We also eliminate galaxies identified as satellites in the redMaPPer cluster catalogue \citep{Ryk++14}. In particular, we reject objects with a cluster membership probability larger than $50\%$ and a central galaxy probability smaller than $10\%$. This step eliminates $\sim5\%$ of the objects.
Finally, we remove extreme outliers by fitting a mass-size and mass-velocity dispersion relation to the sample, and removing objects at more than $4\sigma$ away from the mean relation, with an iterative sigma-clipping procedure.
These outliers could bias the inference, and most likely correspond to faulty measurements in at least one of the three variables considered: mass, size or velocity dispersion. 

Within the resulting sample, $\sim2,000$ galaxies lie in the region of the sky with weak lensing measurements from HSC.
We assign background sources to each lens in the sample. For each lens, we consider sources within a projected distance of $300\,\rm{kpc}$ at the lens redshift. For some lenses, the area within $300\,\rm{kpc}$ overlaps with that of other objects.
In order to avoid using the same source twice, not allowed under the assumption of isolated lenses, we eliminate lenses from the sample until the areas within $300\,\rm{kpc}$ of the remaining lenses no longer overlap. We do this iteratively, starting from the objects with the smallest observed stellar mass. Around 15\% of the lenses are eliminated with this procedure, leaving a final sample of $\sim1,700$ galaxies. These are massive ($\log{\mchab} > 11$) and quiescent galaxies, belonging to the SDSS main spectroscopic sample, with weak lensing measurements from HSC. 
Throughout this paper we refer to this sample of galaxies as the ``control sample''.

The control sample has been obtained by applying cuts to the SDSS main spectroscopic sample, motivated by the need for a sample of massive galaxies with a small contamination from satellites, and clear of more massive neighbours within the region used for the weak lensing analysis.
For consistency, we apply the same criteria to the SLACS sample. This reduces the number of strong lenses available for the analysis to 45 objects, with respect to the initial 59. Of the 14 rejected lenses, 1 is eliminated exclusively for its low stellar mass ($\log{\mchab} < 11.0$), while 13 have more massive galaxies in their proximity. Four of these lenses are also identified as satellites by redMaPPer.

\begin{figure} 
 \includegraphics[width=\columnwidth]{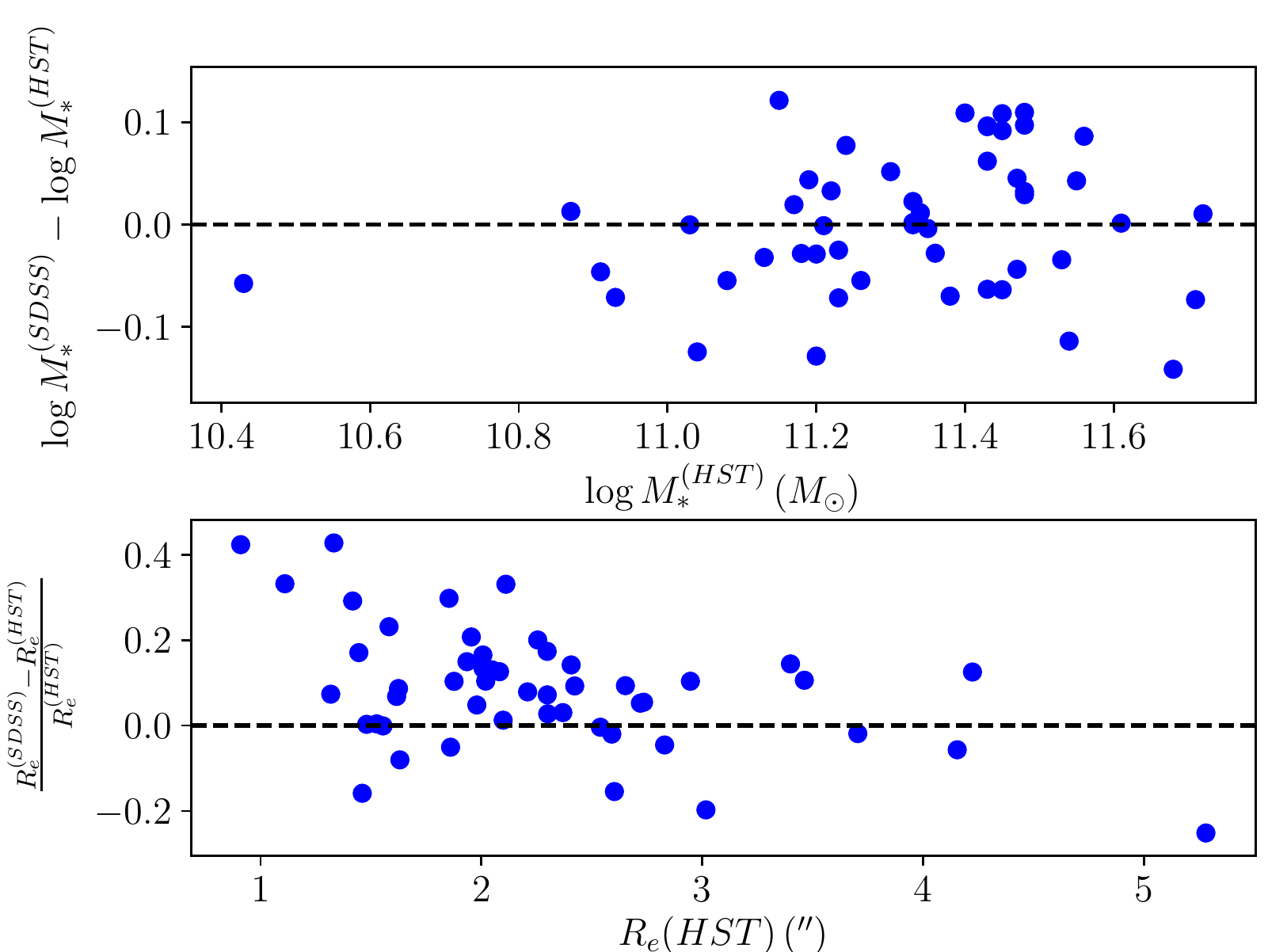}
 \caption{{\em Top:} difference between the SDSS-based stellar mass of SLACS lenses, from MPA-JHU, as a function of the HST-based measurements on the same objects by \citet{Aug++09}.
{\em Bottom:} relative difference between the observed $r$-band effective radii from SDSS DR12 as a function of rest-frame I-band effective radii measured with HST data by \citet{Aug++09}.
\label{fig:mass_size}
}
\end{figure}

Since we will be combining data from this sample with higher quality measurements from the SLACS sample, it is important to make sure that there are no systematic biases between the two datasets.
In particular we want to check if stellar mass and size measurements obtained using ground based data produce the same values as the measurements obtained from space.
This is a relatively straightforward test, since SLACS lenses are in SDSS.
In \Fref{fig:mass_size} we show SDSS stellar masses and effective radii of SLACS lenses as a function of the same quantities measured using HST data.
There is good agreement between the two datasets, particularly in terms of stellar mass.
Effective radii derived from SDSS are on average 8\% larger than HST-based measurements, with a 14\% scatter. Since these are strong lenses, though, it is possible that bias and scatter are not fully representative of the typical galaxies in the control sample, due to the presence of a lensed source blended with the main galaxy that could affect the measurement of $\reff$. In any case, we will discuss the implications of such a small discrepancy at the end of subsection \ref{sect:vanilla}.

\begin{figure} 
 \includegraphics[width=\columnwidth]{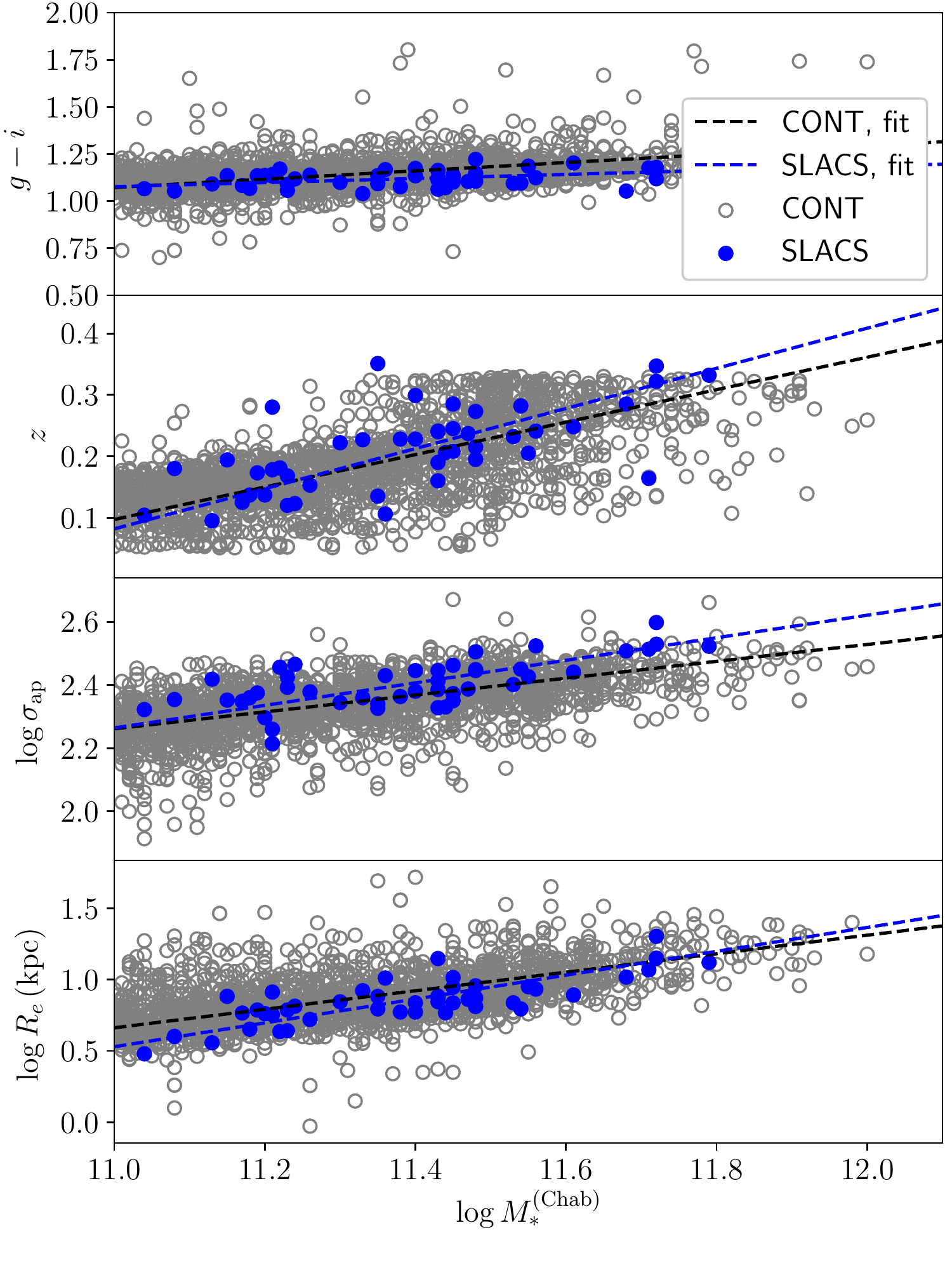}
 \caption{Observed stellar mass vs. effective radius, aperture velocity dispersion, redshift and rest-frame $g-i$ colour, for SLACS lenses and galaxies in the control sample.
The dashed lines show the best fit relation obtained for the two samples.
\label{fig:mstarreffsigma}
}
\end{figure}

Finally, we show in 
\Fref{fig:mstarreffsigma} the distribution in observed stellar mass, effective radius, velocity dispersion, redshift and SDSS rest-frame $g-i$ colour of the SLACS and the control sample.
SLACS galaxies appear to be slightly more compact compared to the control sample, at the low-mass end of the distribution. Their velocity dispersion is correspondingly higher.
The distribution of the two samples in the $M_*-z$ plane is similar, with more massive galaxies being preferentially at higher redshifts, a result of the luminosity selection of the SDSS spectroscopic sample, from which both lens sets are drawn. Given these tests, throughout the rest of this paper, we will assume that the SLACS sample and the control sample are drawn from the same population, modulo the strong lensing selection which will be described in subsection \ref{ssec:lenscorr}.


\subsection{HSC weak lensing and photo-z}

We use weak lensing measurements from the first internal data release (DR1) of the HSC survey shear catalogue \citep{Man++18}. The catalogue covers an area of 137~deg$^2$, with a number density of sources of $25$~arcmin$^{-2}$ (unweighted). Shape measurements have been obtained on $i$-band images using the re-Gaussianization PSF correction method of \citet{H+S03}.
Calibration for multiplicative and additive bias have been obtained through a set of dedicated simulations, similar to those of the GREAT 3 challenge \citep{Man++14, Man++15}. We refer to \citet{Man++18} for further details.

Photometric redshifts for the background sources are taken from the HSC DR1 \citep{Tan++18}. In particular, we use photo-zs obtained through the template fitting-based code {\sc Mizuki} \citep{Tan15}. For each lens, we select sources for which the lower $2\sigma$ bound on the photo-z is larger than the lens redshift. In other words, we select sources with a 97.5\% probability of being in the background of our lenses.
For each source, we take the median value of the photo-z probability distribution as the fiducial value for the source redshift. For simplicity, we do not propagate redshift uncertainties to the shear estimate, as this is sub-dominant with respect to the shape noise \citep[see][for a discussion]{S+L18}.


\subsection{Stacked weak lensing}\label{ssec:stack}

\begin{figure}
 \includegraphics[width=\columnwidth]{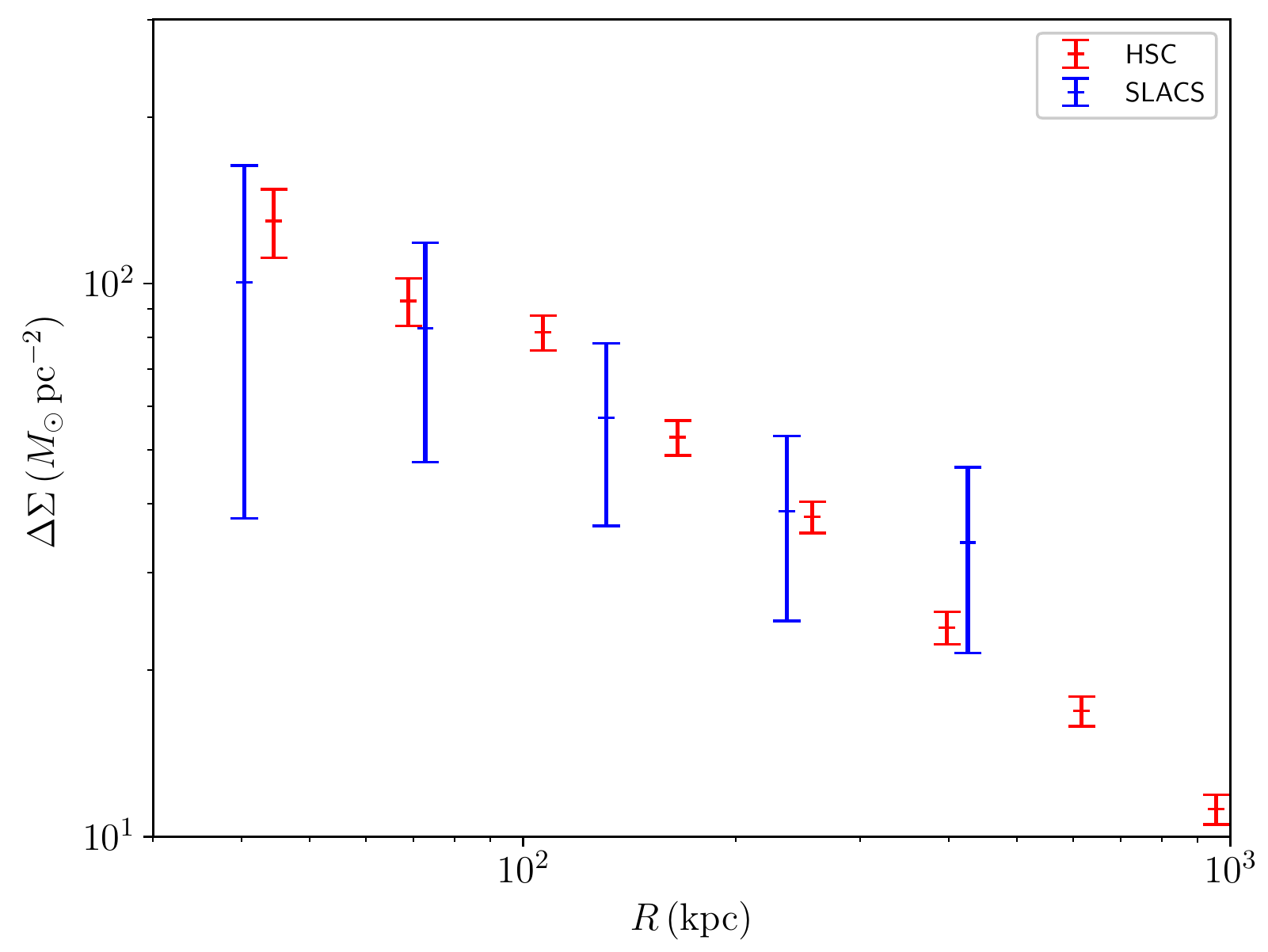}
\caption{Stacked weak lensing signal around SLACS lenses and for our control sample of $1,700$ massive quiescent galaxies with HSC weak lensing. The control sample is matched to the SLACS sample by using a stellar mass-based weighting scheme.
 \label{fig:stacked}
}
\end{figure}

Before introducing the full model used in our analysis we carry out a simple comparison between the stacked weak lensing signal for SLACS lenses and the control sample. The stacked signal for SLACS lenses is obtained by combining, in each radial bin, the measurements of $\Delta\Sigma$ for all the objects in the sample, and converting them to our fiducial cosmology. For the control sample, we combine the signal for all of the $\sim1,700$ lenses, but we apply a weighting scheme to match the stellar mass distribution of the HSC galaxies to that of the SLACS sample.
In particular, for a given galaxy with observed stellar mass $\mchab$, we apply the following weight:
\begin{equation}
w_i = \frac{1}{\sqrt{2\pi}\sigma_{*,\mathrm{(SLACS)}}}\exp{\left\{-\frac{(\log{\mchab} - \mu_{*,\mathrm{(SLACS)}})^2}{2\sigma_{*,\mathrm{(SLACS)}}^2}\right\}},
\end{equation}
where $\mu_{*,\mathrm{(SLACS)}}$ and $\sigma_{*,\mathrm{(SLACS)}}$ are the mean and standard deviation of the observed log-stellar mass of the SLACS sample.
We use the HSC weak lensing pipeline (More et al. in prep.) to calculate $\Delta\Sigma$ in radial bins.
The stacked lensing signal for the SLACS and the control sample is shown in \Fref{fig:stacked}.
There is good agreement between the two samples. Since the stacked weak lensing signal is mainly sensitive to the halo mass distribution, this indicates that SLACS lenses have similar halo masses as non lens galaxies with the same stellar mass. However, because of the small number of SLACS lenses, the SLACS weak lensing measurement is noisy. This prevents us from making more quantitative statements.

\section{Theoretical Framework}\label{sect:model}

We wish to describe the distribution of dark matter halos and stellar IMF at the massive end of the galaxy population.
We do so with a Bayesian hierarchical inference method.
The method takes elements from the analysis of \citet{Son++15}, who applied the Bayesian hierarchical inference formalism to strong lensing and stellar kinematic measurements for 80 galaxies, including SLACS lenses, and the more recent formulation of \citet{S+L18}, who introduced a similar method to interpret weak lensing measurements.

One complication in combining the SLACS sample and the control samples is that the two samples are in principle sampling different regions of the distribution in stellar mass, halo mass and stellar IMF.
Due to strong lensing selection \citep{Ruf++11}, SLACS lenses sample the high mass and high velocity dispersion end of this distribution, as opposed to our sample of non-lenses, which we defined by applying a lower cutoff in stellar mass.
In order to obtain an accurate inference on the halo mass and stellar IMF distribution, it is essential to 1) choose a model that accurately describes how these quantity vary across the region of parameter space spanned by the two samples, and 2) accurately model the strong lensing selection, in order to compare the results inferred from the two samples.

We explore different models. In the first model, labelled the `vanilla' model, we assume that dark matter halos are described by an NFW profile and that halo mass and the normalisation of the stellar IMF scale with a power-law dependence on stellar mass.
This model is described below.
We then generalise the vanilla model in two different ways. The first variation consists in modifying the inner density profile of the dark matter halo by allowing for adiabatic contraction or expansion.
The second variation instead consists in allowing for a gradient in the stellar mass-to-light ratio, while keeping the dark matter halo profile fixed to an NFW profile.
The adiabatic contraction model and the $M_*/L$ gradient model will be described in subsections \ref{sect:adcontr} and \ref{sect:mlgrad} respectively.

\subsection{Bayesian hierarchical inference}\label{ssec:form}

Let us first introduce the concept of Bayesian hierarchical inference in its most basic form.
Let us consider a sample of galaxies, each one fully described by the value of a parameter set $\individ$.
Let us assume that the individual values of the parameters $\individ$ are drawn from a distribution described by a set of hyper-parameters, $\hyperp$, the values of which we wish to infer.
For a given prior, $\pr(\hyperp)$, the posterior probability distribution function (PDF) of the hyper-parameters given a set of measurements $\data$ on the galaxy sample is given by Bayes theorem as
\begin{equation}\label{eq:genbayes}
\pr(\hyperp | \data) \propto \pr(\hyperp) \pr(\data | \hyperp),
\end{equation}
where $\pr(\data | \hyperp)$ is the likelihood of observing the data given the values of the hyper-parameters.
If measurements on different galaxies are independent from each other, this can be expanded as
\begin{equation}\label{eq:genprod}
\pr(\data | \hyperp) = \prod_i \pr(\datai | \hyperp).
\end{equation}
This is not true in general for weak lensing measurements. In principle, each background source is lensed by every foreground galaxy.
\Eref{eq:genprod}, however, is valid under the approximation of isolated lenses, on which our analysis is based. We treat lens galaxies as if they were at infinite distance from each other, so that the likelihood of a set of shape measurements around a given lens is independent of the model describing other lenses. 

Finally, each term in the product can be evaluated by marginalizing over the individual parameters $\individ$ as
\begin{equation}\label{eq:genint}
\pr(\datai | \hyperp) = \int d\individi \pr(\datai | \individi) \pr(\individi | \hyperp).
\end{equation}
Equations (\ref{eq:genbayes}), (\ref{eq:genprod}) and (\ref{eq:genint}) allow us to make an inference on the values of the hyper-parameters $\hyperp$ by evaluating the posterior PDF $\pr(\hyperp | \data)$.
There is a two-level hierarchy of parameters in this problem: the hyper-parameters $\hyperp$, at the top level, specify the distribution of lower-level parameters $\individ$, describing individual objects, which are needed to calculate the likelihood.

\subsection{Individual galaxy parameters}

We now apply this formalism to our specific science case.
We describe the mass distribution of each galaxy as the sum of two components. A spherically de-projected de Vaucouleurs profile describing the stars, and a spherical NFW profile for the dark matter halo.
We then introduce the IMF mismatch parameter, defined as the ratio between the true stellar mass of a galaxy, and the stellar mass an observer would infer assuming a Chabrier IMF and having otherwise perfect knowledge of the stellar population parameters of the galaxy, $\mchab$:
\begin{equation}
\aimf \equiv \frac{\mstar}{\mchab}.
\end{equation}
We use the following parameters to describe each galaxy:
\begin{equation}\label{eq:individ}
\individ \equiv \{\mchab, \aimf, \mhalo, \chalo\},
\end{equation}
where $\mhalo$ is the halo mass and $\chalo$ the halo concentration.

\subsection{Hyper-parameters of the galaxy population distribution}

At this point we need to specify a form for the distribution of galaxy parameters, $\individ$ described by a given set of hyper-parameters.
We assume the following form:
\begin{equation}\label{eq:dist}
\begin{split}
\pr(\individ | \hyperp) = & \mathcal{S}(\mchab)\mathcal{H}(\mhalo | \mchab)\times \\
& \mathcal{C}(\chalo | \mhalo)\mathcal{I}(\aimf | \mchab). 
\end{split}
\end{equation}
The term $\mathcal{S}(\mchab)$ describes the distribution in stellar mass. Following \citet{S+L18} we use a skew Gaussian distribution, which is shown to be appropriate for the description of a sample obtained by applying a stellar mass cut:
\begin{equation}\label{eq:fullskew}
\begin{split}
\mathcal{S}(\mchab) = & \frac{1}{\sqrt{2\pi}\sigma_*}\exp{\left\{-\frac{(\log{\mchab} - \mu_*)^2}{2\sigma_*^2}\right\}}\times \\
& \Phi(\log{\mchab}),
\end{split}
\end{equation}
with
\begin{equation}\label{eq:skew}
\Phi(\log{\mchab}) = 1 + \mathrm{erf}\left(s_*\frac{\log{\mchab} - \mu_*}{\sqrt{2}\sigma_*}\right).
\end{equation}
The values of the parameters $\mu_*$, $\sigma_*$ and $s_*$ will naturally be different for the two samples of galaxies, the SLACS lenses and the control sample.

The next term in \Eref{eq:dist} is $\mathcal{H}(\mhalo | \mchab)$, describing the distribution of halo masses as a function of stellar mass, also modelled as a Gaussian:
\begin{equation}\label{eq:haloterm}
\mathcal{H}(\mhalo | \mchab) = \frac{1}{\sqrt{2\pi}\sigma_h}\exp{\left\{-\frac{(\log{\mhalo} - \mu_h)^2}{2\sigma_h^2}\right\}}.
\end{equation}
As anticipated, we allow the mean of this Gaussian to be a function of stellar mass:
\begin{equation}\label{eq:muhalo}
\mu_h = \mu_{h,0} + \beta_h(\log{\mchab} - 11.3).
\end{equation}
The halo mass-concentration relation is modelled as
\begin{equation}\label{eq:massconc}
\pr(\chalo | \mhalo) = \frac{1}{\sqrt{2\pi}\sigma_c}\exp{\left\{-\frac{(\log{\chalo} - \mu_c(\mhalo))^2}{2\sigma_c^2}\right\}}
\end{equation}
with
\begin{equation}\label{eq:maccio}
\mu_c(\mhalo) = c_{h, 0} + \beta_c(\log{\mhalo} - 12).
\end{equation}
We fix the values of the parameters of the mass-concentration relation to $c_{h, 0} = 0.830$, $\beta_c=-0.098$ and $\sigma_c=0.1$, based on results from numerical simulations by \citet{Mac++08}.

Finally we introduce a distribution for $\alpha_{\rm IMF}$:
\begin{equation}
\mathcal{I}(\aimf | \mchab) = \frac{1}{\sqrt{2\pi}\sigma_{\mathrm{IMF}}}\exp{\left\{-\frac{(\log{\aimf} - \mu_{\mathrm{IMF}})^2}{2\sigma_{\mathrm{IMF}}^2}\right\}}.
\end{equation}
Similarly to the halo mass distribution, we allow the mean of this Gaussian to scale with stellar mass
\begin{equation}
\mu_{\mathrm{IMF}}= \mu_{\mathrm{IMF},0} + \beta_{\mathrm{IMF}}(\log{\mchab} - 11.3).
\end{equation}
We ignore redshift evolution in the model parameters, since the redshift range spanned by the lens samples is quite small. 
In summary, the full list of hyper-parameters is
\begin{equation}\label{eq:hplist}
\hyperp \equiv \{\mu_*, \sigma_*, s_*, \mu_{h,0}, \beta_h, \mu_{\mathrm{IMF},0}, \beta_{\mathrm{IMF}}, \sigma_{\mathrm{IMF}}\}.
\end{equation}
A brief description of each parameter is provided in \Tref{tab:vanilla}.

\subsection{The likelihood term}

In order to evaluate the posterior PDF of the hyper-parameters we need to calculate the likelihood term $\pr(\data | \individ)$ for each galaxy.
The data consist of:
\begin{itemize}
\item Observed stellar mass from stellar population synthesis, $\mobs$.
\item Line-of-sight velocity dispersion measured within the SDSS spectroscopic aperture, $\sigmaap$.
\item Weak lensing shape measurements, $\wl$ 
\item Einstein radius $\rein$, defined as the radius of a circular aperture within which the mean surface mass density equals the critical density of the lens-source pair (SLACS lenses only).
\end{itemize}
The likelihood is then given by
\begin{equation}
\begin{split}
\pr(\data | \individ) = & \pr(\mobs | \mchab)\pr(\sigmaap | \mchab, \aimf, \mhalo, \chalo) \times \\
& \pr(\wl | \mchab, \aimf, \mhalo, \chalo) \times \\ 
& \pr(\rein | \mchab, \aimf, \mhalo, \chalo).
\end{split}
\end{equation}
The stellar mass term $\pr(\mobs | \mchab)$ is a Gaussian in $\mchab$. The velocity dispersion term is evaluated using the spherical Jeans equation to construct a model seeing-convolved luminosity-weighted line of sight velocity dispersion within the spectroscopic aperture.
The strong and weak lensing terms are calculated by producing model mass density profiles composed of the sum of a de Vaucouleurs profile of mass $\aimf\mchab$ for the stars and a dark matter halo, assuming that the two have the same centre and no other mass component contributes to the lensing signal in the region probed by the data.

The likelihood for HSC weak lensing is given by the following product over individual lensed sources:
\begin{multline}
\pr(\wl | \mchab, \aimf, \mhalo, \chalo) = \\
\prod_i \pr(e_{t,i}|\mchab, \aimf, \mhalo, \chalo).
\end{multline}

Here, $e_{t,i}$ indicates the tangential component of the observed ellipticity of the $i$-th source, relative to the lens. Unlike ellipticity measurements based on other methods, the expectation value of $e_{t,i}$ defined with the re-Gaussianization method of \citet{H+S03} is not equal to the tangential component of the shear, $g_t$. Instead, the following identity holds:
\begin{equation}\label{eq:regauss}
2R g_t = \left<e_t\right>,
\end{equation}
where $R$ is called the `shear responsivity', $R\approx 1 - e_{\mathrm{rms}}^2$, and $e_{\mathrm{rms}}$ is the intrinsic RMS scatter in source ellipticities.
\Eref{eq:regauss} is valid for unbiased ellipticity estimates. In practice, the expectation value of the ellipticity is corrected with a multiplicative and additive bias, as follows:
\begin{equation}
2R[(1 + m)g_t + c_t] = \left<e_t\right>,
\end{equation}
where $c_t$ is the tangential component of the additive bias. $m$ and $c$ are calibrated on simulations, as described by \citet{Man++18}.
The likelihood of observing a given value of the tangential ellipticity given the model is then
\begin{multline}
\pr(e_t | \mchab, \aimf, \mhalo, \chalo) = \\
\frac{1}{\sqrt{2\pi\sigma_{e,\mathrm{tot}}^2}}\exp{\left\{-\frac{(2R[(1+m)g_t + c_t] - e_t)^2}{2\sigma_{e,\mathrm{tot}}^2}\right\}},
\end{multline}
where $\sigma_{e,\mathrm{tot}}^2$ is the sum in quadrature of the intrinsic shape noise $e_{\mathrm{rms}}$ and the shape measurement error, $\sigma_e$.
The reduced shear $g_t$ is a function of the model parameters, as well as the redshifts of lens and sources.

\subsection{Goodness of fit estimation}\label{ssec:pp}

We now discuss how the goodness of fit of our model is assessed. In Bayesian hierarchical inference methods, this is usually done through {\em posterior predictive tests}: we use the posterior probability distribution to generate mock data, which we then compare with the observed data.
In principle, the goodness of fit can then be determined by defining a metric quantifying how well the model prediction matches the observations. For simplicity, however, we will limit our evaluation to a qualitative comparison.

 In our case, the relevant data consists of weak lensing shape measurements, Einstein radii and central velocity dispersions.
For both galaxy samples, we will attempt to reproduce their stacked weak lensing signal.
For the strong lens sample, we will also predict Einstein radii and stellar velocity dispersion measurements. We will first compare predicted and observed values of $\rein$ and $\sigmaap$ separately, and then we will combine these to obtain a measurement of the slope of the total density profile, $\gamma'$. The total density slope has been measured for lenses in the SLACS sample by fitting power-law density profiles, $\rho(r) \propto r^{-\gamma'}$, to the Einstein radius and velocity dispersion measurements \citep{Koo++06, Aug++10}. We will use the same procedure to determine posterior predicted values of $\gamma'$. 

The density slope has a more intuitive physical meaning compared to the Einstein radius. Whereas $\rein$ indicates the total mass enclosed within an aperture of arbitrary radius, determined by the geometry of the lens-source system, the total density slope $\gamma'$ provides information on the internal structure of a lens.
Although the value $\gamma'$ is not a pure observable, and depends somewhat on the assumption of a power-law density profile \citep[see][for a detailed study]{Xu++17}, its use should be seen as a way of mapping observables from the two-dimensional space defined by $\rein$ and $\sigmaap$ to a one-dimensional variable. At fixed $\sigmaap$, increasing $\rein$ corresponds to a smaller value of $\gamma'$ (a shallower density profile). At fixed $\rein$, increasing $\sigmaap$ results in a larger $\gamma'$ (a steeper density profile).

In practice, we proceed as follows. We first draw a subset of points from the MCMC sample of the posterior probability distribution for a given model. For each point, corresponding to a set of hyper-parameter values, we generate a mock lens population, with values of stellar mass, stellar IMF, halo mass and concentration drawn from the model distribution. Additionally, we draw values for the lens redshift distribution, which we model as a Gaussian centred at $0.18$ with $0.07$ dispersion, which is a good approximation for both samples. Finally, we draw values for effective radii from a mass-size relation. This is modelled as a Gaussian in $\reff$ with mean given by

\begin{equation}
\mu_{R} = \mu_{R,0} + \beta_R(\log{\mchab} - 11.3),
\end{equation}
with $\mu_{R,0}=0.85$, $\beta_R=0.65$, and dispersion $\sigma_R=0.13$. These values are determined by fitting the mass-size relation of the control sample.

Given a mock lens population, we compute the average $\Delta\Sigma$ at the location of the radial bins used in \Fref{fig:stacked}. For consistency with subsection \ref{ssec:stack}, we apply the same stellar mass-based weighting scheme to $\Delta\Sigma$ for the mock based on the inference from the control sample.

For the SLACS-based mock, we also generate a population of background sources. We draw source redshifts from a Gaussian distribution centred at $0.60$ and with a dispersion of $0.18$, which is a good approximation of the redshift distribution of SLACS strongly-lensed sources.
We then compute Einstein radii and, by solving the spherical Jeans equation, the seeing-convolved surface brightness-weighted line-of-sight velocity dispersion within a circular aperture of $1.5''$ radius. Finally, we fit a power-law density profile to the mock Einstein radius and velocity dispersion data to obtain the density slope $\gamma'$.

We then compare the distribution in $\Delta\Sigma$, $\rein$, $\sigmaap$ and $\gamma'$ obtained from mocks generated using 1,000 randomly drawn points from the MCMC chain with the observed stacked weak lensing and density slope measurements.

\subsection{The Strong Lensing Selection Function}\label{ssec:lenscorr}

In order to compare the inferences from both galaxy samples, strong lensing selection effects need to be accounted for.
There are two separate selections at play. The first is due to different galaxies having different cross sections for strong lensing. This is relatively easy to model, as will be shown later.
The second is due to some strong lenses being easier to discover than others, depending on their properties and on the efficiency of the lens search algorithm.

\begin{figure}
 \includegraphics[width=\columnwidth]{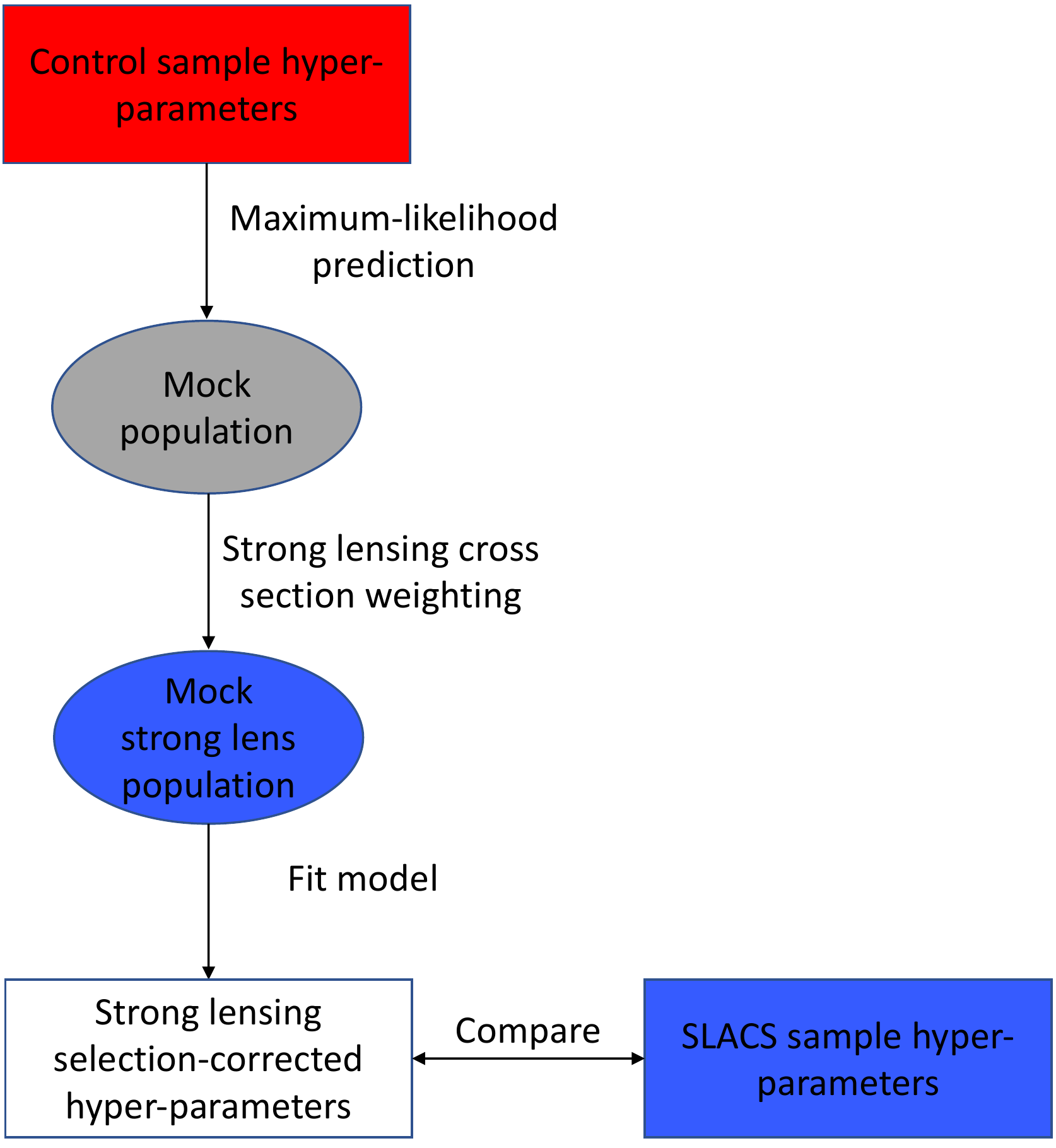}
\caption{Schematic representation of the strategy adopted to compare the inferences from the SLACS sample and the control sample, while taking into account strong lensing selection effects.
\label{fig:scheme}
}
\end{figure}

In principle, strong lensing selection can be explicitly modelled in our Bayesian formalism. This would be achieved by describing the distribution of SLACS galaxies as the product of the distribution of the control sample times a term proportional to the product between the lensing cross section and the lens finding efficiency, which skews it towards the position in parameter space occupied by strong lenses. This approach was used in the strong lensing and stellar dynamics analysis of \citet{Son++15}.
However, the relatively high dimensionality of our model makes it technically challenging to implement this approach here. Instead, we carry out a posterior predictive test.
We start from the maximum likelihood model inferred from the control sample, which we assume to be the truth, and use it to draw a mock population. We populate these galaxies with background sources with the same redshift distribution as that of SLACS lensed sources and calculate the strong lensing cross section and lens finding efficiency of each lens-source pair. We then draw a random subsample of galaxies, weighting each galaxy by its strong lensing cross section and lens finding efficiency. This subsample is meant to simulate a sample of strong lenses.
Finally, we fit our model to this subsample.
The inferred hyper-parameters will represent the distribution of SLACS galaxies as predicted from the control sample. A schematic representation of the process is shown in \Fref{fig:scheme}.

Strictly speaking, the procedure we just described is not a posterior predictive test because we only focus on the maximum-likelihood set of hyper-parameters. In principle, one should perform a posterior predictive measurement at each point of the hyper-parameter space sampled by the control sample posterior probability distribution, and obtain a new distribution of hyper-parameters to be compared with the SLACS inference.
In practice, this would require too much computational time. We then only consider how the lensing selection shifts the peak of the posterior, and assume that the same shift applies to the whole distribution.

We define the strong lens cross-section as the area in the source plane that gets mapped into sets of at least two multiple images with a minimum magnification of $|\mu_{min}| = 0.5$. This last criterion is chosen in order to exclude lens configurations with very de-magnified images: although in principle these are still strong lens systems, in practice they cannot be identified as such.
We assume circular symmetry for the cross-section computation.
We verified that the results are stable with respect to variations in the value of $\mu_{min}$.

The lens finding efficiency is more difficult to model.
SLACS lenses were found by means of a spectroscopic search for emission lines from lensed background sources, using the $1.5''$~radius SDSS fibre. 
Qualitatively, we understand how the lens finding efficiency varies as a function of the Einstein radius: it goes to zero for very small values of $\rein$, since the flux from the lensed source is correspondingly very small, peaks for values of $\rein$ close to the fibre radius, and goes down for very large values of $\rein$, since most of the flux falls outside of the spectroscopic fibre for these lensed sources.
However, a quantitative assessment of the lens finding efficiency requires dedicated simulations, including the creation of synthetic spectra, the exploration of different mock lens configurations, and modelling the effects of the atmospheric seeing. 

\citet{ABB12} carried out such a study in the context of power-law lens models. 
Their main result is that, for a SLACS-like survey, the lens finding efficiency is zero for very small values of $\rein$, is approximately constant for values of the Einstein radius up to $\rein\simeq2.0''$, then falls with increasing $\rein$ for lenses with a density profile steeper than isothermal \citep[see the middle left panel of Figure 5 of][]{ABB12}.
With this result in mind, we approximate the lens finding efficiency as a constant for values of $\rein$ down to $0.5''$, and zero for smaller values.
This is a good approximation, since most of the Einstein radii of the population of mock strong lenses have values $\rein < 2.0''$.
We verified that none of our results change when the lens finding efficiency is set to zero for $\rein > 2.0''$.


\section{Results}\label{sect:results}

\subsection{NFW halos and spatially constant IMF}\label{sect:vanilla}

We fit the model described in \Sref{sect:model} to the observed stellar masses, velocity dispersions, background source shapes.  For SLACS galaxies we also use Einstein radii. The SLACS sample and the control sample are fit separately. The posterior PDF is evaluated by running a Markov Chain Monte Carlo. We assume uniform priors on all hyper-parameters except for $s_*$. For $s_*$ we assume a uniform prior on its logarithm in the range $(-1, 1)$. Integrals over individual lens parameters in \Eref{eq:genint} are computed with Monte Carlo integration and importance sampling, following \citet{S+L18} \citep[see also][]{Sch++15}.

\begin{figure*}
 \includegraphics[width=\textwidth]{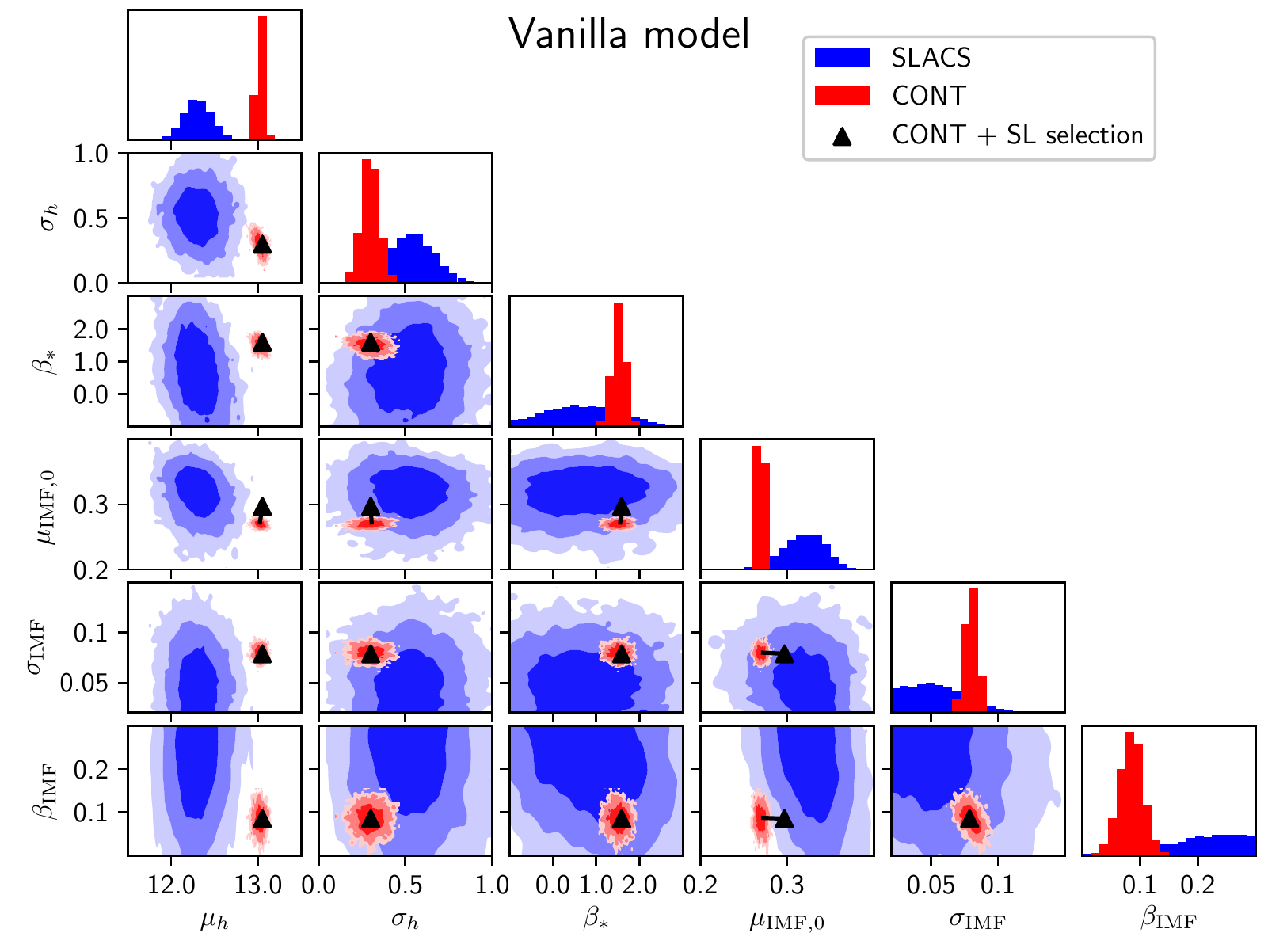}
\caption{Vanilla model. Posterior probability distribution of the parameters describing the distribution in halo mass and IMF mismatch parameter. {\em Red region:} inference from the control sample. {\em Blue lines:} inference from SLACS galaxies.
{\em Black triangles:} values of the hyper-parameters obtained by applying a strong lensing selection correction to the maximum-likelihood control sample model, as described in subsection \ref{ssec:lenscorr}
 \label{fig:vanillacp}
}
\end{figure*}

\Fref{fig:vanillacp} displays the posterior PDF relative to the hyper-parameters describing the halo mass and stellar IMF distribution, inferred from the SLACS and control samples separately.
Median values with 68\% confidence region of the inference on the full list of hyper-parameters are reported in \Tref{tab:vanilla}.
The values of the hyper-parameters resulting from applying the strong lensing selection correction to the maximum-likelihood model for the control sample, as described in subsection \ref{ssec:lenscorr}, are plotted as black triangles in \Fref{fig:vanillacp}, and listed in \Tref{tab:vanilla}.
The shifts with respect to the maximum-likelihood model due to lensing selection are minimal. The largest effect is seen for the average IMF normalisation parameter, $\mu_{\mathrm{IMF},0}$, which increases by $\sim0.05$ and is brought into better agreement with the SLACS inference.
However, lensing selection has a very small effect on halo mass. 
The reason for this is that the projected dark matter mass within the Einstein radius of galaxies, typically on the scale of $5$~kpc, is a very mild function of halo mass.
\begin{table*}
\caption{Vanilla model. Median values and 68\% confidence interval of the posterior probability distribution of individual hyper-parameters, marginalised over the rest of the hyper-parameters. The third column lists values of the hyper-parameters obtained by applying a strong lensing selection correction to a model population corresponding to the maximum-likelihood of the control sample inference, as described in subsection \ref{ssec:lenscorr}.
In the last row we list the inferred median stellar mass of the different samples.
}
\label{tab:vanilla}
\begin{tabular}{lcccl}
\hline
\hline
 & SLACS & Control & Control, SL pred. & Parameter description \\
\hline
$\mu_{h,0}$ & $12.31 \pm 0.17$ & $13.02 \pm 0.04$ & 13.05 & Average $\log{\mhalo}$ at stellar mass $\log{\mchab}=11.3$\\
$\sigma_h$ & $0.52 \pm 0.15$ & $0.30 \pm 0.05$ & 0.31 & Dispersion in $\log{\mhalo}$ around the average\\
$\beta_h$ & $0.64 \pm 0.91$ & $1.51 \pm 0.15$ & 1.56 & Power-law dependence of halo mass on $\mchab$\\
$\mu_{\mathrm{IMF}, 0}$ & $0.32 \pm 0.03$ & $0.27 \pm 0.01$ & 0.30 & Average $\log{\aimf}$ at stellar mass $\log{\mchab}=11.3$\\
$\sigma_{\mathrm{IMF}}$ & $0.05 \pm 0.02$ & $0.08 \pm 0.01$ & 0.08 & Dispersion in $\log{\aimf}$ around the average\\
$\beta_{\mathrm{IMF}}$ & $0.29 \pm 0.13$ & $0.08 \pm 0.02$ & 0.08 & Power-law dependence of IMF normalization on $\mchab$\\
$\mu_*$ & $11.29 \pm 0.09$ & $11.23 \pm 0.03$ & 11.38 & Average parameter in \Eref{eq:fullskew}\\
$\sigma_*$ & $0.22 \pm 0.05$ & $0.23 \pm 0.02$ & 0.22 & Dispersion parameter in \Eref{eq:fullskew}\\
$\log{s_*}$ & $-0.54 \pm 0.85$ & $-0.21 \pm 0.19$ & -0.89 & Log of the skewness parameter in \Eref{eq:skew}\\
$\rm{Median}\,\log{M_*}$ & $11.37 \pm 0.03$ & $11.33 \pm 0.01$ & 11.41 & Median stellar mass (not a hyper-parameter)\\

\hline
\end{tabular}
\end{table*}

Let us first consider the inference on the hyper-parameters obtained from the control sample. 
The derived average in $\log{\mhalo}$ at the pivot stellar mass $\log{\mchab} = 11.3$ is $\mu_{h,0} = 13.02 \pm 0.04$, while the average of the distribution in $\log{\aimf}$ is $\mu_{\mathrm{IMF},0} = 0.27 \pm 0.01$. This result is consistent with the stacked weak lensing measurement of \citet{SMP10}, who concluded that a stellar IMF consistent with a Salpeter IMF (which would correspond to $\log{\aimf}=0.25$) is needed to match the dynamical masses (i.e. the velocity dispersion measurements) of a sample of SDSS galaxies selected in a similar way to our control sample.

The inference based on the SLACS sample is for the most part consistent with that based on the control sample, with one important exception: halo mass.
SLACS data favours a very low value for the average halo mass: $\mu_{h,0} = 12.31 \pm 0.17$.
This value is not only inconsistent with the inference from HSC data, but, together with the inferred value of the IMF normalisation $\mu_{\mathrm{IMF},0}=0.32$, would correspond to a baryonic fraction of 15\% of the total mass, very close to the cosmological value of $\sim16\%$ \citep{Planck++16}, and therefore unphysical.

This result might seem in contradiction with the plot of \Fref{fig:stacked}, in which the stacked weak lensing signal measured for SLACS lenses is observed to be consistent with that of the control sample, once matched in stellar mass.
However, when fitting our model to the whole set of available data for the SLACS lenses, the likelihood is mostly determined by the constraints on small scales, strong lensing and stellar kinematics, as the precision on the Einstein radius and velocity dispersion measurements is much higher compared to the weak lensing measurements.
The low value of the inferred average halo mass, then, indicates that the model is not a good description of the structure of massive galaxies at all the scales probed by the data.

\begin{figure}
\includegraphics[width=\columnwidth]{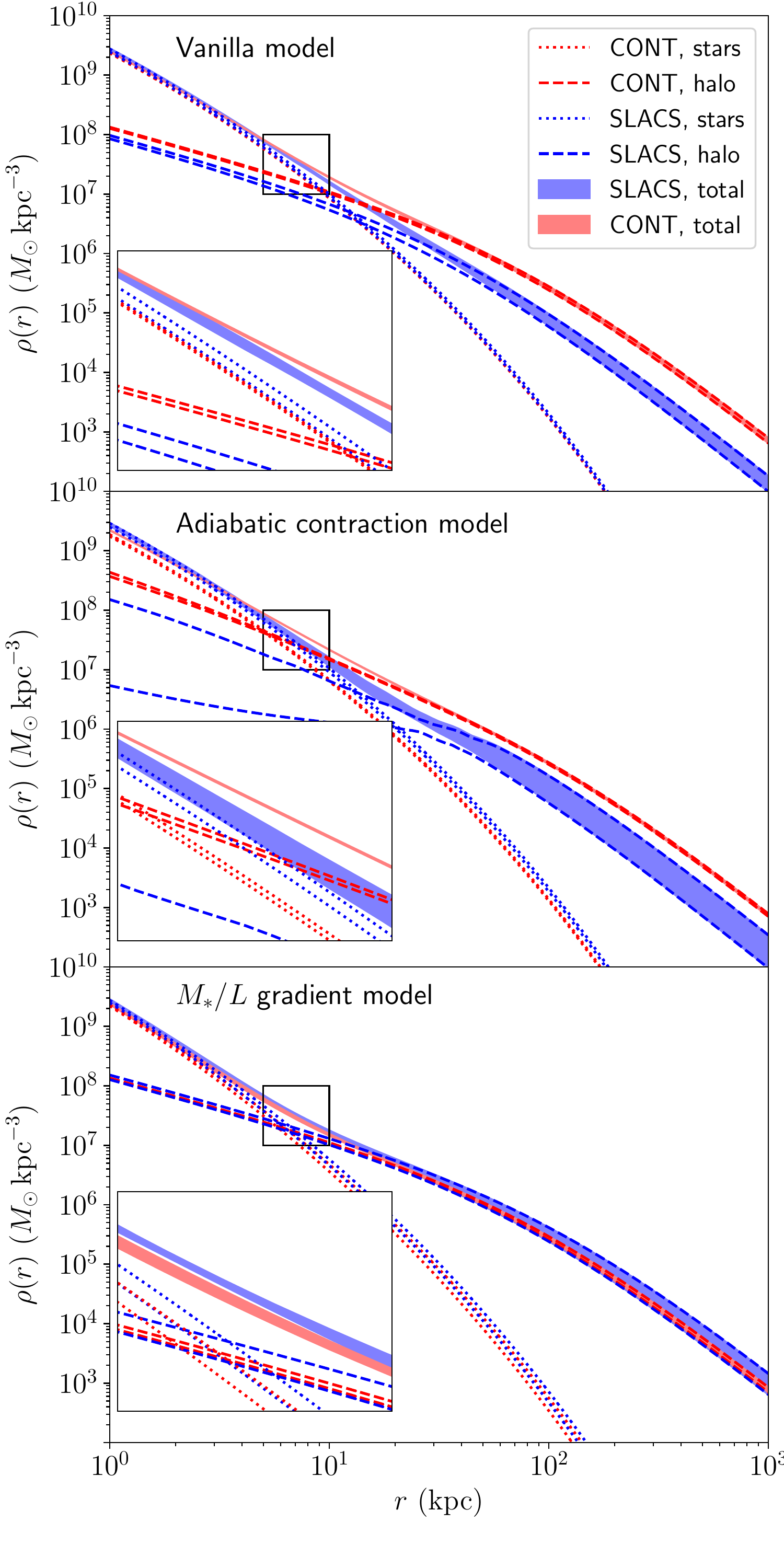}
\caption{Average density profile of a galaxy with $\log{\mchab}=11.3$ and $\reff=7\,\rm{kpc}$ inferred by fitting the vanilla (top), adiabatic contraction (middle) and $M_*/L$ gradient (bottom) model to the SLACS (blue) and control sample (red) datasets. {\em Shaded region:} 68\% confidence region on the total density profile. {\em Dashed lines:} 84\% and 16\% enclosed probability limits on the dark matter density. {\em Dotted lines:} 84\% and 16\% enclosed probability limits on the stellar mass density.
In each panel, the inset shows in greater detail the region of the plot delimited by the rectangle. 
\label{fig:rho_merged}
}
\end{figure}

The discrepancy between the inferences for the two datasets is further illustrated in the top panel of \Fref{fig:rho_merged}, where we plot the average stellar, dark matter, and total density profile inferred for a galaxy of stellar mass $\log{\mchab} = 11.3$ and effective radius $\reff=7\,\rm{kpc}$.
The inner density profile is quite similar in the two cases. This is because, for both datasets, the model is fit to the SDSS velocity dispersion, which is mostly sensitive to the total density profile on the scales of a few kpc.
On the other hand, the two density profiles diverge drastically at large radii, due to the large difference in the inferred halo mass.

Although in principle, for this comparison, we should be considering the average density profile obtained after applying the strong lensing selection correction to the control sample, the differences with respect to the uncorrected profile are minimal, and we do not plot it in \Fref{fig:rho_merged} to minimise confusion.

\begin{figure}
 \includegraphics[width=\columnwidth]{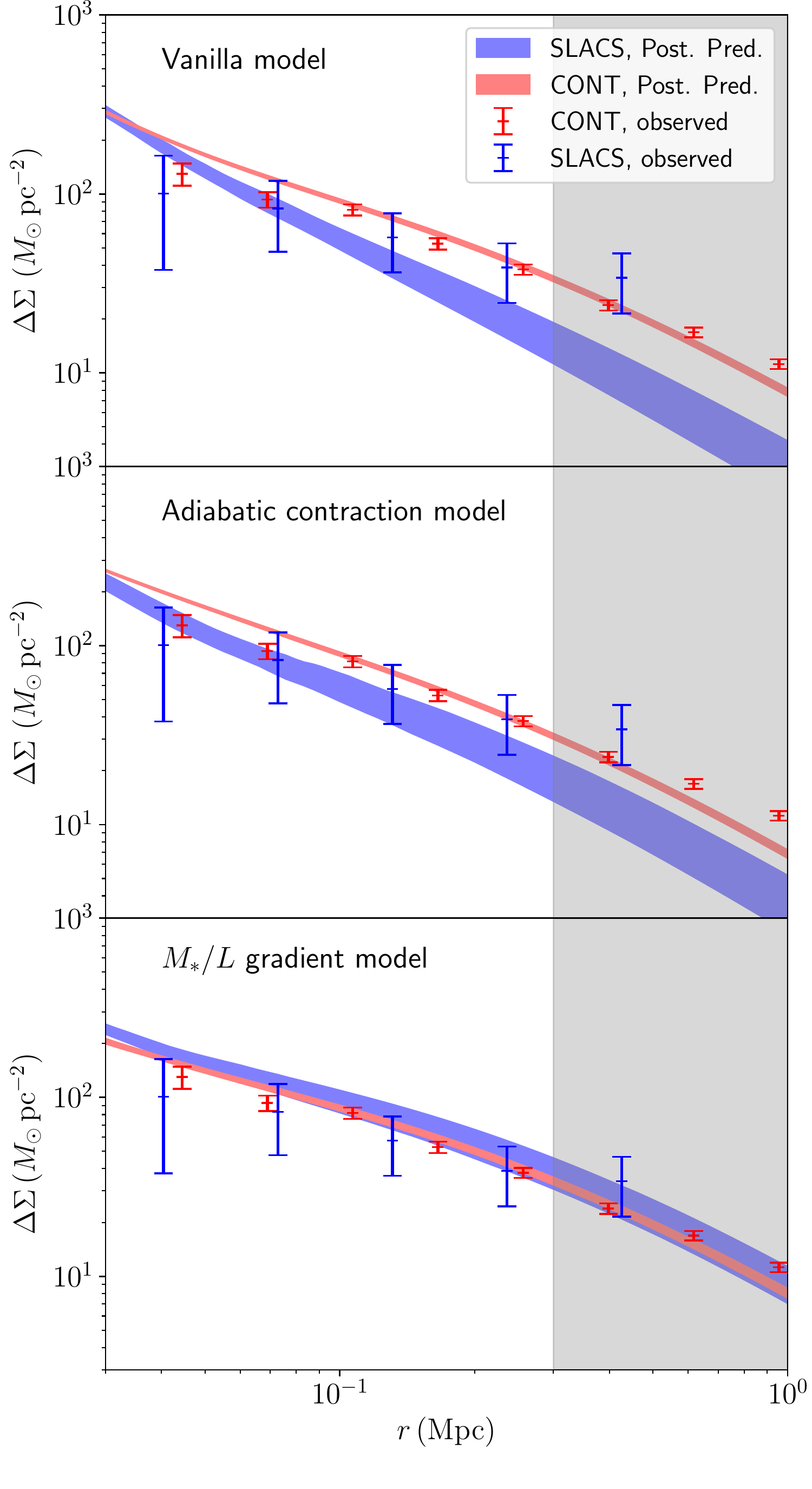}
 \caption{Posterior predictive tests based on the stacked weak lensing profile, to assess the goodness of fit of the vanilla (top), adiabatic contraction (middle) and $M_*/L$ gradient (bottom) model. Observed data points from \Fref{fig:stacked} are shown as error bars. The shaded bands correspond to regions of 68\% enclosed probability in the average $\Delta\Sigma$. In the case of the control sample, both the posterior predictive and the observed distribution are obtained by applying a stellar mass weighting scheme based on the stellar mass distribution of SLACS lenses.
The shaded region indicates the radial range not used to constrain the model.
\label{fig:deltasigma_pp}
}
\end{figure}

Let us now consider the goodness of fit. In the top panel of \Fref{fig:deltasigma_pp}, we plot the posterior predictive stacked weak lensing signal, generated as described in subsection \ref{ssec:pp}, compared with the observed data.
For the control sample, the model is a good description of the data.
The same can be said for the model based on the SLACS sample: although the inferred halo mass is very low, the predicted $\Delta\Sigma$ profile goes through the data points, in virtue of the large uncertainties of the SLACS weak lensing measurements.

\begin{figure*}
 \includegraphics[width=\textwidth]{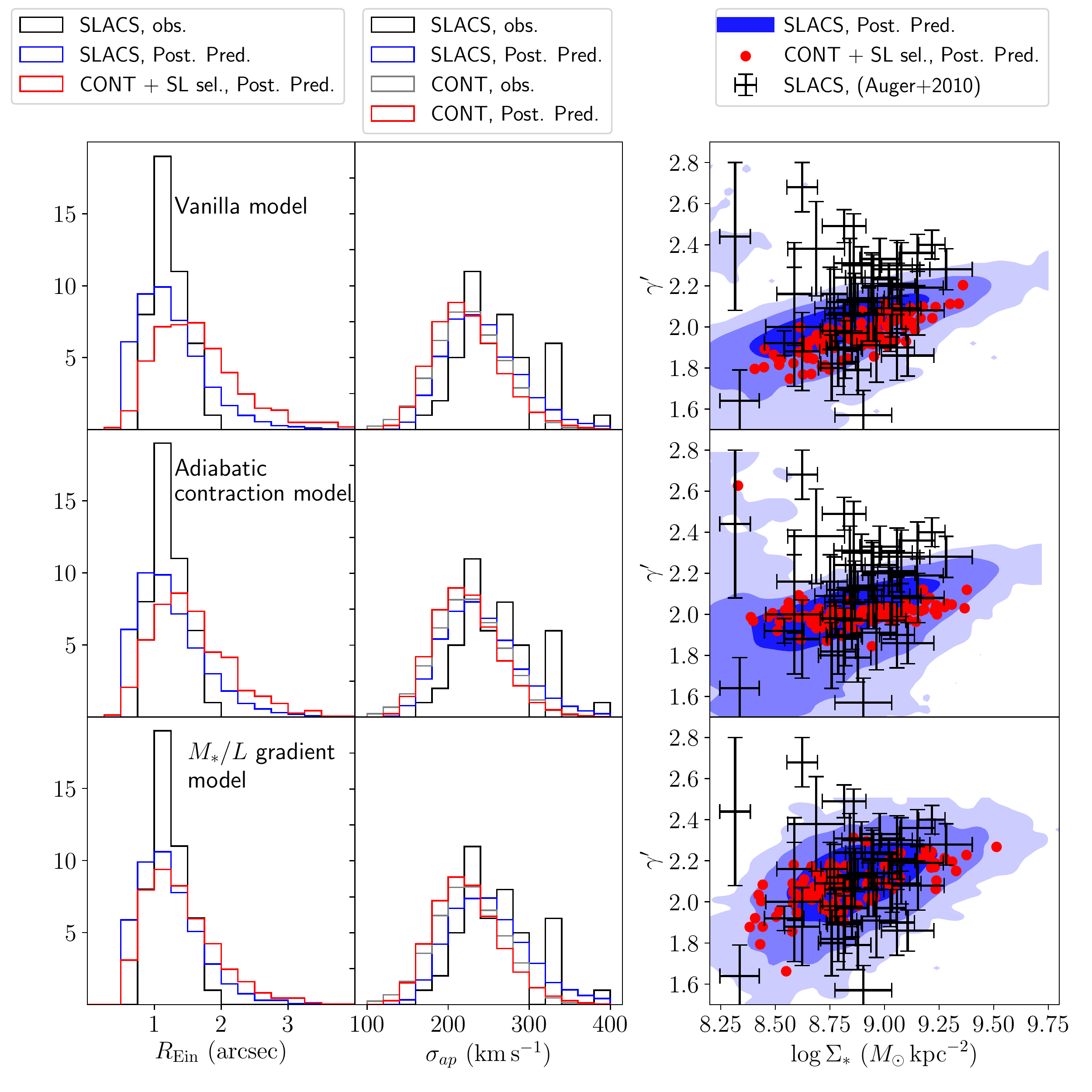}
 \caption{Posterior predictive test on strong lensing and stellar kinematics observables, to assess the goodness of fit of the vanilla (top), adiabatic contraction (middle) and $M_*/L$ gradient (bottom) model. {\em Left panel}: distribution in Einstein radius. The black histogram is the observed distribution from \citet{Aug++09}, the blue one is obtained from the posterior inferred using SLACS data, the red one is obtained from the peak of the posterior obtained from the control sample, corrected by strong lensing selection (i.e. weighted by strong lensing cross section). Histograms are normalised to the total number of SLACS lenses. {\em Middle panel}: distribution in SDSS velocity dispersion. Different colours correspond to the observed distribution of SLACS lenses from \citet{Aug++09} (black), the observed distribution of the control sample (grey), and the posterior predicted distributions based on the SLACS (blue) and control sample (red) inferences. {\em Right panel}: total density slope, obtained by fitting a power-law density profile to $\rein$ and $\sigmaap$, as a function of stellar mass density. Black error bars correspond to values obtained by \citet{Aug++10} on the SLACS sample. The shaded region is the distribution obtained by drawing samples of galaxies from the posterior given by the SLACS inference. Red circles are samples drawn from the peak of the control sample posterior, corrected by strong lensing selection.
\label{fig:rein_sigma_gammap}
}
\end{figure*}

\Fref{fig:rein_sigma_gammap} focuses on small-scale observables: the Einstein radius, the velocity dispersion and the inner slope of the total density profile, $\gamma'$, derived from the combination of the two as described in subsection \ref{ssec:pp}. 
As can be seen from the first two plots in the top row, the SLACS-based model produces a reasonable distribution in Einstein radius and velocity dispersion (blue histograms), compared to the observed distributions (black histograms).

The corresponding density slope $\gamma'$ is plotted in the upper right panel as a function of stellar mass density $\Sigma_* = \mchab/(2\pi\reff^2)$, which has been shown to correlate with $\gamma'$ \citep{Son++13b}.
The posterior predictive distribution of $\gamma'$ based on the SLACS sample seems to reproduce reasonably well the observed distribution, including the trend with $\Sigma_*$.
In this respect, we recover the result of \citet{Sha++17}, who showed that the correlation of the density slope with stellar mass and size can be explained with NFW + de Vaucouleurs models and a simple stellar-to-halo mass relation.
However, there are a few SLACS lenses that cannot be described by the posterior predictive $\gamma'$ distribution. These are objects with $\log{\Sigma_*}\approx 9.0$ and large values of the density slope $\gamma' \approx 2.3$.
The vanilla model is unable to reproduce such steep density profiles. 

This is particularly the case for the values of the hyper-parameters derived using HSC data (the control sample): the predictive distribution generated from the maximum-likelihood point of the HSC-based inference, corrected for strong lensing selection effects and shown as red circles in the upper right panel of \Fref{fig:rein_sigma_gammap}, covers only the low-$\gamma'$ end of the observed distribution.
This is another way of understanding why a simple NFW + de Vaucouleurs model is not a good description of the data: if the model is tuned to reproduce the weak lensing signal at large radii, the predicted inner density profile is too shallow.
The low values of $\gamma'$ are in turn a result of the predicted Einstein radii being too large compared to the observed distribution, as can be seen by comparing the red and black histograms in the upper left panel of \Fref{fig:rein_sigma_gammap}: for a given value of the velocity dispersion, a larger Einstein radius corresponds to a shallower density profile, and vice-versa.

The statistical errors on the hyper-parameters are significantly smaller for the inference based on the control sample, compared to the SLACS sample. This is especially the case for the hyper-parameter describing the average stellar IMF normalisation, $\mu_{\mathrm{IMF},0}$, for which we obtained a $0.01$~dex uncertainty.
The stellar IMF is constrained, for the control sample, mostly by the SDSS velocity dispersion measurements. This is because the velocity dispersion is mostly sensitive to the mass distribution within the region probed by the stars, which is dominated by the baryons. Dark matter parameters, on the other hand, are mostly constrained by weak lensing data, at larger radii.
Our `vanilla' model has only one degree of freedom in the stellar component: its mass, given by the product $\mchab\aimf$.
The model is then well constrained by the data, given that observational uncertainties on the velocity dispersion are on the order of $10\,\rm{km}\,{s}^{-1}$ ($\sim5\%$ relative uncertainty on $\sigma_{ap}$).
The combination of $1,700$ measurements then brings the statistical error on the mean IMF normalisation to a very small value.
However, the true uncertainty on the IMF normalisation parameter is most likely dominated by systematic uncertainties related to our particular model choice. This will become evident in subsection \ref{sect:mlgrad}.

One possible source of systematic uncertainty is the measurement of the effective radius. We have shown in \Fref{fig:mass_size} how the SDSS-based values of $\reff$, which we use for our control sample, appear to be systematically larger by 8\% compared to HST-based measurements on the same galaxies. Assuming this offset to be real, and not due to contamination from the lensed background source (we used SLACS lenses for this comparison), we can estimate the impact on the inference on $\aimf$ with a simple dynamical argument.
The IMF normalisation that we derive from the control sample is essentially determined by the ratio between a dynamical mass, obtained from the velocity dispersion measurements, and a stellar mass obtained from photometry. From the virial theorem, the dynamical mass scales with the effective radius and the velocity dispersion as follows
\begin{equation}
M_{dyn} \sim \frac{\reff \sigma^2}{G}.
\end{equation}
Therefore, errors on $\reff$ propagate linearly to the dynamical mass, and, to first approximation, to the IMF normalisation. An 8\% systematic error on the effective radius then corresponds to a similar error on $\aimf$.
We therefore assume a systematic uncertainty of $\sim0.03$ on the hyper-parameter $\mu_{\mathrm{IMF},0}$, to be added in quadrature to the statistical uncertainty.

\subsection{Modified dark matter profile}\label{sect:adcontr}

The vanilla model produces an unphysical solution when fitted to SLACS data.
This prompts us to propose a different model.
We then relax the assumption of an NFW profile for the dark matter halo and allow for the effect of adiabatic contraction or expansion.
We define adiabatically contracted/expanded halos following \citet{Dut++07}.
For a given halo mass, stellar mass and effective radius, a fully contracted halo profile is calculated using the adiabatic contraction prescription of \citet{Blu++86}. The \citet{Blu++86} treatment assumes that baryons and dark matter share initially the same density profile, which we assume to be an NFW halo with a standard mass-concentration relation. As the baryons contract to the observed distribution (which we describe with a spherically de-projected de Vaucouleurs profile) the orbits of dark matter particles are modified while conserving the product $r M(r)$, where $r$ is the distance from the centre of the potential and $M(r)$ is the total mass enclosed within the spherical shell of radius $r$. The product $r M(r)$ is an adiabatic invariant in the case of circular orbits.

Starting from the fully contracted dark matter profile we then build a family of profiles by introducing the contraction efficiency parameter $\nu$, which is allowed to range between $1$, corresponding to a fully contracted profile \`{a} la \citet{Blu++86} and $-1$, corresponding to an expanded profile.

\begin{figure} 
\includegraphics[width=\columnwidth]{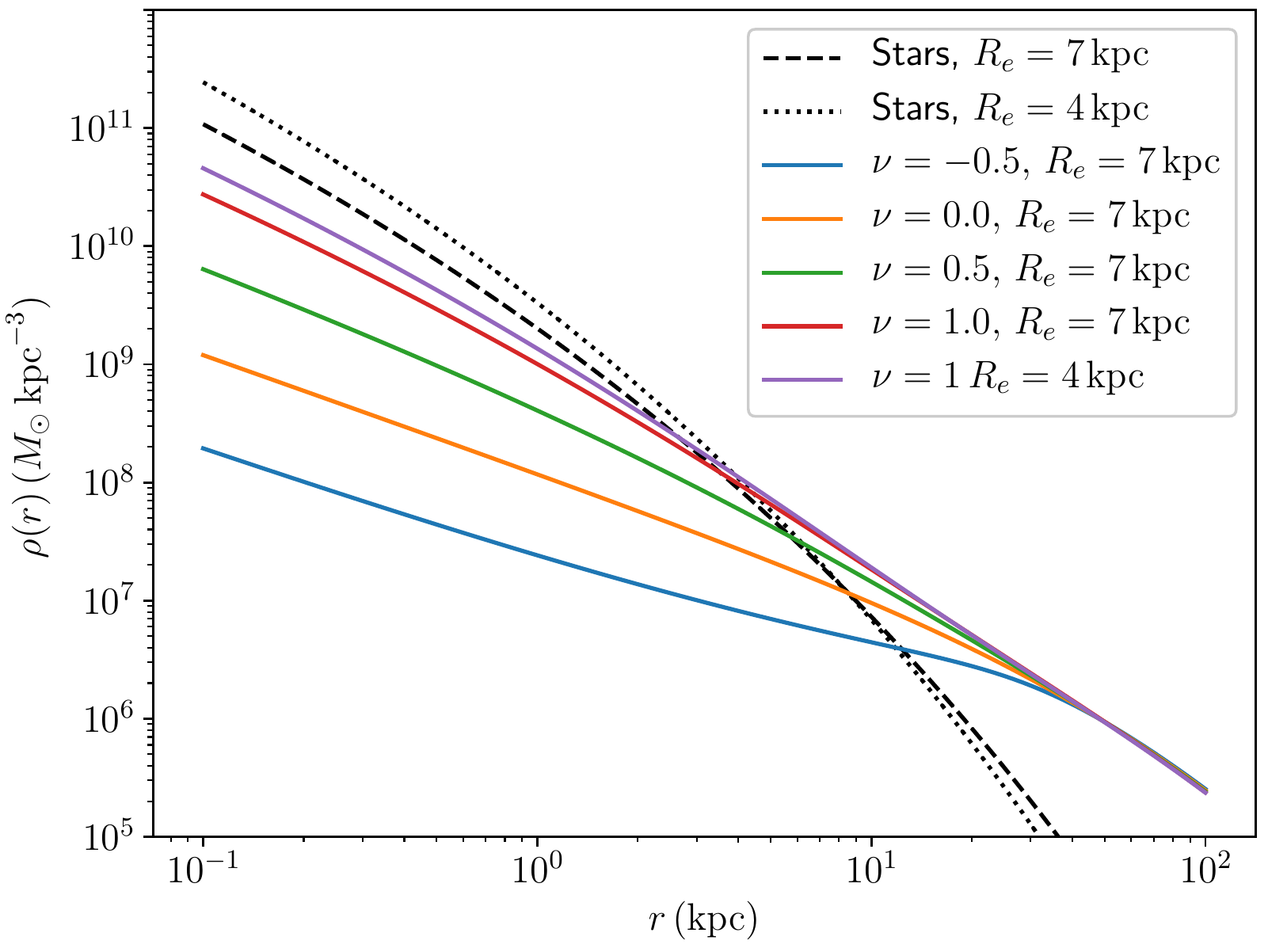}
 \caption{Density profile of adiabatically contracted/expanded halo for different values of the response efficiency parameter $\nu$. For $\nu=0$, the profile is that of an NFW halo. In this example the halo mass is $\log{\mhalo}=13$, the halo concentration is $\chalo=5$, the stellar mass is $\log{M_*}=11.5$ and the effective radius is $7\,\rm{kpc}$ or $4\,\rm{kpc}$. The density profile of the baryonic component for the two effective radii is also plotted.
\label{fig:acexample}
}
\end{figure}

The resulting dark matter density profile for different values of the parameter $\nu$ and for two different values of the galaxy effective radius is plotted in \Fref{fig:acexample}.
The effects of adiabatic contraction or expansion are visible only in the inner $\sim50\,\rm{kpc}$.
We also point out how, for the same halo mass, stellar mass and contraction efficiency parameter $\nu$, the effect of adiabatic contraction is stronger for more compact galaxies. 

We model the distribution of this new parameter, $\nu$, as a Gaussian, truncated between $-1 < \nu < 1$, which we multiply to \Eref{eq:dist}. The new probability distribution of the individual galaxy parameters given the hyper-parameters becomes
\begin{equation}\label{eq:acdist}
\begin{split}
\pr(\individ | \hyperp) = & \mathcal{S}(\mchab)\mathcal{H}(\mhalo | \mchab)\times \\
& \mathcal{C}(\chalo | \mhalo)\mathcal{I}(\aimf | \mchab) \mathcal{N}(\nu), 
\end{split}
\end{equation}
with
\begin{equation}
\mathcal{N}(\nu) = \frac{A_\nu}{\sqrt{2\pi}\sigma_\nu}\exp{\left\{-\frac{(\nu - \mu_\nu)^2}{2\sigma_\nu^2}\right\}};\quad -1 < \nu < 1.
\end{equation}
The new hyper-parameters $\mu_\nu$ and $\sigma_\nu$ are the mean and dispersion of this truncated Gaussian, on which the normalisation constant $A_\nu$ depends.
We assume the following exponential prior on the parameter $\sigma_\nu$
\begin{equation}
\pr(\sigma_\nu) \propto \exp{\left(-\frac{\sigma_\nu}{0.1}\right)},
\end{equation}
in order to penalize solutions with very large values of the intrinsic scatter in the adiabatic contraction efficiency, which would correspond to a flat distribution in $\nu$. In other words, we assert a model in which the distribution in the adiabatic contraction efficiency is centered around a well-defined mean.

\begin{figure*}
 \includegraphics[width=\textwidth]{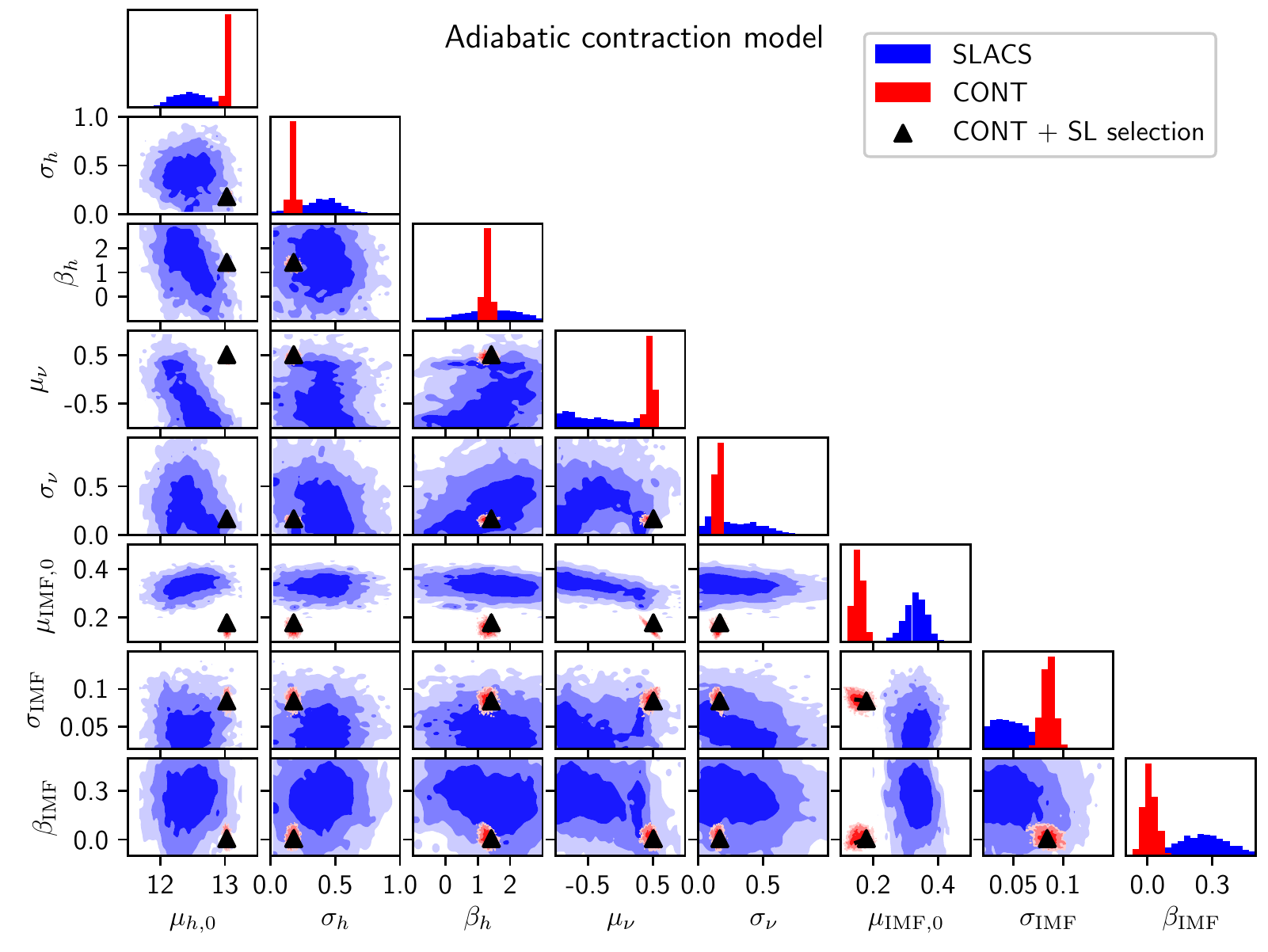}
\caption{Adiabatic contraction model. Posterior probability distribution of the parameters describing the distribution in halo mass, IMF mismatch parameter and adiabatic contraction efficiency. {\em Red region:} inference from the control sample. {\em Blue lines:} inference from SLACS lenses.
{\em Black triangles:} values of the hyper-parameters obtained by applying a strong lensing selection correction to the maximum-likelihood control sample model, as described in subsection \ref{ssec:lenscorr}.
 \label{fig:adcontrcp}
}
\end{figure*}

We fit this new model to SLACS data and the control sample separately. The inference on the hyper-parameters describing halo mass, stellar IMF and adiabatic contraction efficiency is plotted in \Fref{fig:adcontrcp}, while median and 68\% confidence interval of all hyper-parameters are listed in \Tref{tab:adcontr}.
\begin{table*}
\caption{Adiabatic contraction model. Median values and 68\% confidence interval of the posterior probability distribution of individual hyper-parameters, marginalised over the rest of the hyper-parameters. The third column lists values of the hyper-parameters obtained by applying a strong lensing selection correction to a model population corresponding to the maximum-likelihood of the control sample inference, as described in subsection \ref{ssec:lenscorr}.
}
\label{tab:adcontr}
\begin{tabular}{lcccl}
\hline
\hline
 & SLACS & Control & Control, SL pred. & Parameter description \\
\hline
$\mu_{h,0}$ & $12.45 \pm 0.25$ & $13.03 \pm 0.03$ & 13.03 & Average $\log{\mhalo}$ at stellar mass $\log{\mchab}=11.3$\\
$\sigma_h$ & $0.39 \pm 0.16$ & $0.17 \pm 0.02$ & 0.18 & Dispersion in $\log{\mhalo}$ around the average\\
$\beta_h$ & $1.28 \pm 0.92$ & $1.29 \pm 0.10$ & 1.42 & Power-law dependence of halo mass on $\mchab$\\
$\mu_\nu$ & $-0.38 \pm 0.44$ & $0.47 \pm 0.05$ & 0.51 & Average adiabatic contraction efficiency parameter\\
$\sigma_\nu$ & $0.31 \pm 0.20$ & $0.15 \pm 0.02$ & 0.17 & Dispersion in adiabatic contraction efficiency parameter\\
$\mu_{\mathrm{IMF}, 0}$ & $0.33 \pm 0.03$ & $0.15 \pm 0.02$ & 0.18 & Average $\log{\aimf}$ at stellar mass $\log{\mchab}=11.3$\\
$\sigma_{\mathrm{IMF}}$ & $0.05 \pm 0.02$ & $0.09 \pm 0.01$ & 0.08 & Dispersion in $\log{\aimf}$ around the average\\
$\beta_{\mathrm{IMF}}$ & $0.26 \pm 0.14$ & $0.01 \pm 0.03$ & 0.01 & Power-law dependence of IMF normalization on $\mchab$\\
$\mu_*$ & $11.30 \pm 0.09$ & $11.25 \pm 0.03$ & 11.34 & Average parameter in \Eref{eq:fullskew}\\
$\sigma_*$ & $0.22 \pm 0.05$ & $0.23 \pm 0.01$ & 0.23 & Dispersion parameter in \Eref{eq:fullskew}\\
$\log{s_*}$ & $-0.58 \pm 0.82$ & $-0.32 \pm 0.18$ & -0.45 & Log of the skewness parameter in \Eref{eq:skew}\\
$\rm{Median}\,\log{M_*}$ & $11.38 \pm 0.03$ & $11.33 \pm 0.01$ & 11.40 & Median stellar mass (not a hyper-parameter) \\

\hline
\end{tabular}
\end{table*}

Allowing for adiabatic contraction or expansion does help bringing the average halo mass inferred from SLACS lenses in a $2\sigma$ agreement with the control sample inference.
However, a careful look at the $\mu_\nu - \mu_h$ degeneracy contour (first column, fourth row in \Fref{fig:adcontrcp}) shows that strong lensing data allows for reasonably large average halo masses only for negative values of the adiabatic contraction efficiency parameter, corresponding to adiabatic expansion. The control sample, instead, favors slight contraction.
The inference on the IMF normalisation is also correspondingly different: much higher for SLACS galaxies.

The discrepancy between the two inferences can also be seen by looking at the density profile, plotted in the middle panel of \Fref{fig:rho_merged}: although the total density profile of the two models agree, the inferred central dark matter density differs by at least a factor of a few between the two datasets.
Once again, our model is unable to simultaneously describe both datasets with the same values of the hyper-parameters.

The middle panels of \Fref{fig:deltasigma_pp} and \Fref{fig:rein_sigma_gammap} show posterior predictive tests for this model, through which we can check the goodness of fit. Compared to the vanilla model, in this case the model based on SLACS data provides a better description of the stacked weak lensing signal. The adiabatic contraction model, however, is still unable to describe the inner density profile of a subset of SLACS lenses with relatively large values of $\gamma'$, as can be seen from the middle right panel of \Fref{fig:rein_sigma_gammap}.
These objects, when fitted with a bulge + halo model, prefer solutions with very little dark matter \citep[see e.g.][]{Pos++15}.
In fact, even mass-follows-light models for some of these lenses under-predict their measured stellar velocity dispersion.
As a reference, a strong lens consisting only of a de Vaucouleurs profile (no dark matter) corresponds to $\gamma'\approx2.2$ \citet[see red triangles in Figure 6 of][]{Son++13b}, a value exceeded by 16 out of the 45 strong lenses in our sample.

Adiabatically contracted or expanded dark matter profiles do not help improve the match with the data in these cases, because this family of models can only produce dark matter profiles that are shallower than the stellar component, as can be seen from \Fref{fig:acexample}.
As a result, the data favors solutions with as little dark matter in the inner regions as possible, either via a low average halo mass or through adiabatic expansion.
We have verified this conjecture by repeating the fit over a smaller sample of strong lenses, obtained by eliminating the 16 objects with a power-law density slope $\gamma'$ steeper than $2.2$. In this case, we find values of the hyper-parameters that are consistent with those inferred from the control sample, indicating that the problem lies in the objects with a steep density profile.

In principle, we could allow for even greater freedom on the choice of the dark matter profile. For instance, we could also vary the halo concentration, which at the moment is fixed by a relatively narrow prior.
We suspect that models for which the inner slope of the dark matter density profile is steeper than the stellar component would finally allow us to obtain a good fit.
This could be achieved by setting the scale radius of the dark matter halo to be comparable to the Einstein radius, or, in other words, by increasing the concentration by a factor of ten or more.
However, such steep density profiles would be difficult to justify theoretically: any baryonic effect additional to adiabatic contraction would tend to flatten the profile.

\subsection{Gradients in stellar M/L}\label{sect:mlgrad}

In this subsection we introduce a new model, with more freedom in the stellar mass distribution, and with the dark matter model reverted to an NFW profile.
We introduce a new parameter describing a radial gradient in stellar mass-to-light ratio.
While until now we have implicitly assumed that the stellar mass follows exactly the light distribution, here we relax this assumption and allow for the following mass-to-light ratio profile
\begin{equation}\label{eq:grad}
\Upsilon_*(R) = \Upsilon_{*,e} \left(\frac{R}{\reff}\right)^{\gamma},
\end{equation}
where $\Upsilon_{*,e}$ is the mass-to-light ratio at the effective radius, and $-1 < \gamma < 0$.
With this definition, for negative values of $\gamma$ the mass-to-light ratio in the inner regions of the galaxy is larger than in the outskirts. 
Although this parametrization of the $M_*/L$ profile is unphysical at very small and large radii, where it tends to infinity and zero respectively, this is not a problem because the contribution to the total budget of the stellar mass in these regions is very small. In other words, the results would not change if we add an upper and lower limit to the mass-to-light ratio.
However, the above parametrization allows for a relatively easy implementation of a mass-to-light ratio gradient in our analysis.

There can be two origins for $M_*/L$ gradients. One is the presence of gradients in age, dust, or metallicity across the galaxy. Another possibility is the presence of gradients in stellar IMF.
Gradients in stellar population properties at fixed IMF can in principle be measured with spatially resolved spectroscopic data, or estimated with high quality multi-band photometry \citep[see e.g.][]{San++17}.
Our stellar population measurements do not provide us with spatially resolved information.
The stellar population synthesis-based stellar masses used so far, $\mchab$, are obtained fitting synthetic stellar population models to total magnitudes of galaxies.
The resulting stellar mass-to-light ratio is therefore a light-weighted average over the extent of the whole galaxy.
We then define the light-weighted mass-to-light ratio as follows:
\begin{equation}
\left<\Upsilon_*\right>_{\mathrm{LW}} = \frac{\int_0^\infty I(R) R \Upsilon_*(R) dR}{\int_0^\infty I(R) R dR},
\end{equation}
where $I(R)$ is the surface brightness profile of the galaxy, in our case assumed to be a de Vaucouleurs profile.

\begin{figure} 
\includegraphics[width=\columnwidth]{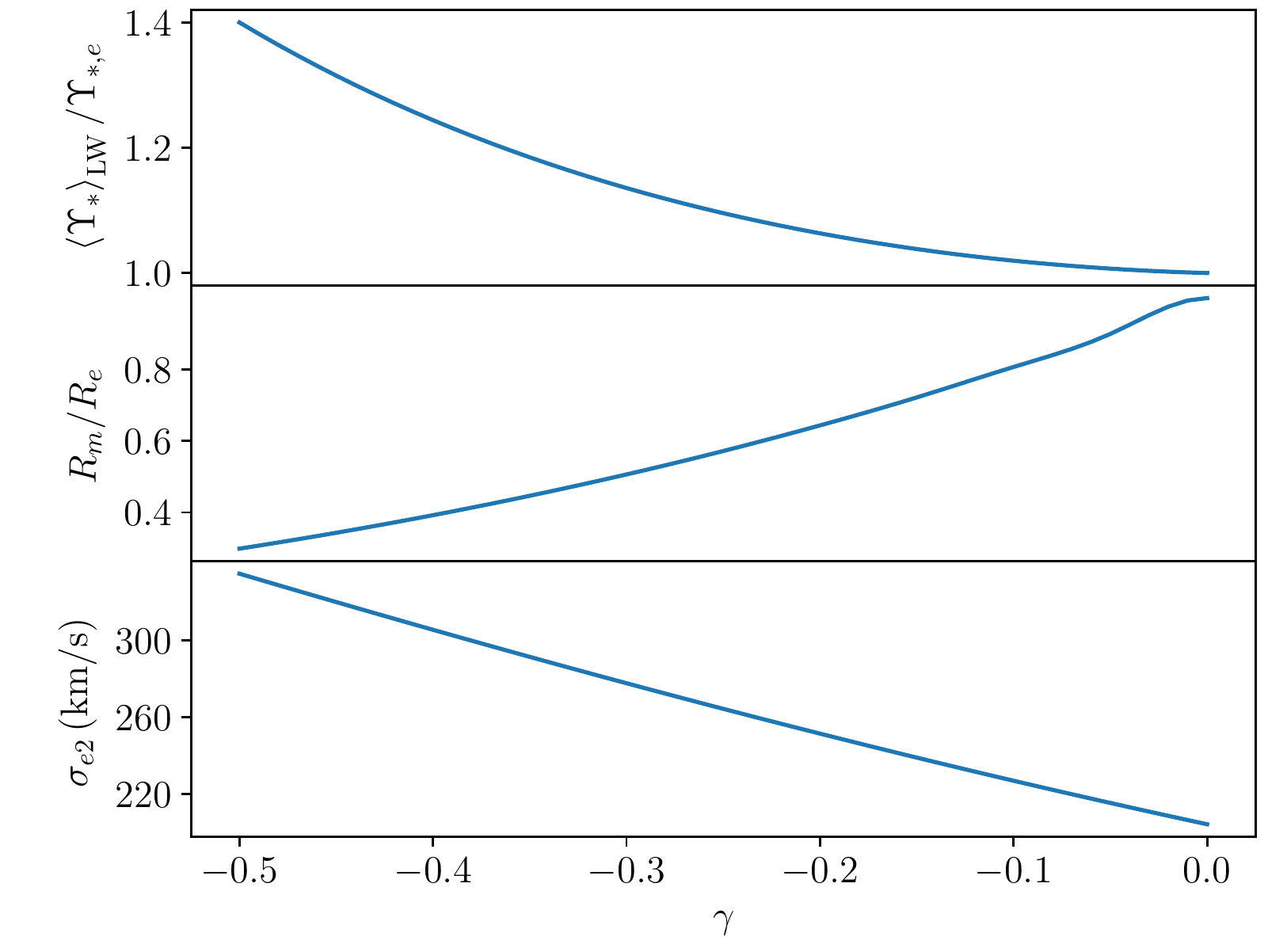}
 \caption{{\em Top:} Ratio between light-weighted stellar mass-to-light ratio and stellar mass-to-light ratio at the effective radius, as a function of the M/L gradient slope $\gamma$.
{\em Center:} Ratio between half-mass radius and half-light radius as a function of gradient slope $\gamma$.
{\em Bottom:} Central velocity dispersion for a galaxy with stellar mass $\log{\mstar}=11.5$, halo mass $\log{\mhalo}=13$ and effective radius $\reff=7\,\rm{kpc}$, as a function of $M_*/L$ radial slope. The model velocity dispersion is calculated by solving the spherical Jeans equation under the assumption of isotropic orbits.
\label{fig:upsilon}
}
\end{figure}

In \Fref{fig:upsilon} we plot the relation between $\Upsilon_{*,e}$ and $\left<\Upsilon_*\right>_{\mathrm{LW}}$.
In addition, we compute the ratio between half-mass radius and half-light radius, as well as the impact of the gradient on the central velocity dispersion, plotted in \Fref{fig:upsilon} as a function of $\gamma$.

This parametrization allows for a steeper total density profile compared to our previous stellar mass-follows-light model.
Strong lensing and stellar kinematics data are particularly sensitive to this new parametrization, since it can modify significantly the total density profile in the inner few kpc. 
As can be seen in \Fref{fig:upsilon}, changing the slope of the $M_*/L$ gradient can change significantly the velocity dispersion of a galaxy \citep[see also][for a related discussion]{Ber++18}.

Similarly to the adiabatic contraction efficiency case, we describe the distribution of the gradient parameter $\gamma$ as a truncated Gaussian of mean $\mu_\gamma$ and dispersion $\sigma_\gamma$.
The new probability distribution of the individual galaxy parameters given the hyper-parameters becomes
\begin{equation}\label{eq:graddist}
\begin{split}
\pr(\individ | \hyperp) = & \mathcal{S}(\mchab)\mathcal{H}(\mhalo | \mchab)\times \\
& \mathcal{C}(\chalo | \mhalo)\mathcal{I}(\aimf | \mchab) \mathcal{G}(\gamma), 
\end{split}
\end{equation}
with
\begin{equation}
\mathcal{G}(\gamma) = \frac{A_\gamma}{\sqrt{2\pi}\sigma_\gamma}\exp{\left\{-\frac{(\gamma - \mu_\gamma)^2}{2\sigma_\gamma^2}\right\}}.
\end{equation}
We assume an exponential prior on the intrinsic scatter in the gradient parameter:
\begin{equation}
\pr(\sigma_\gamma) \propto \exp{\left(-\frac{\sigma_\gamma}{0.1}\right)}.
\end{equation}

\begin{figure*}
 \includegraphics[width=\textwidth]{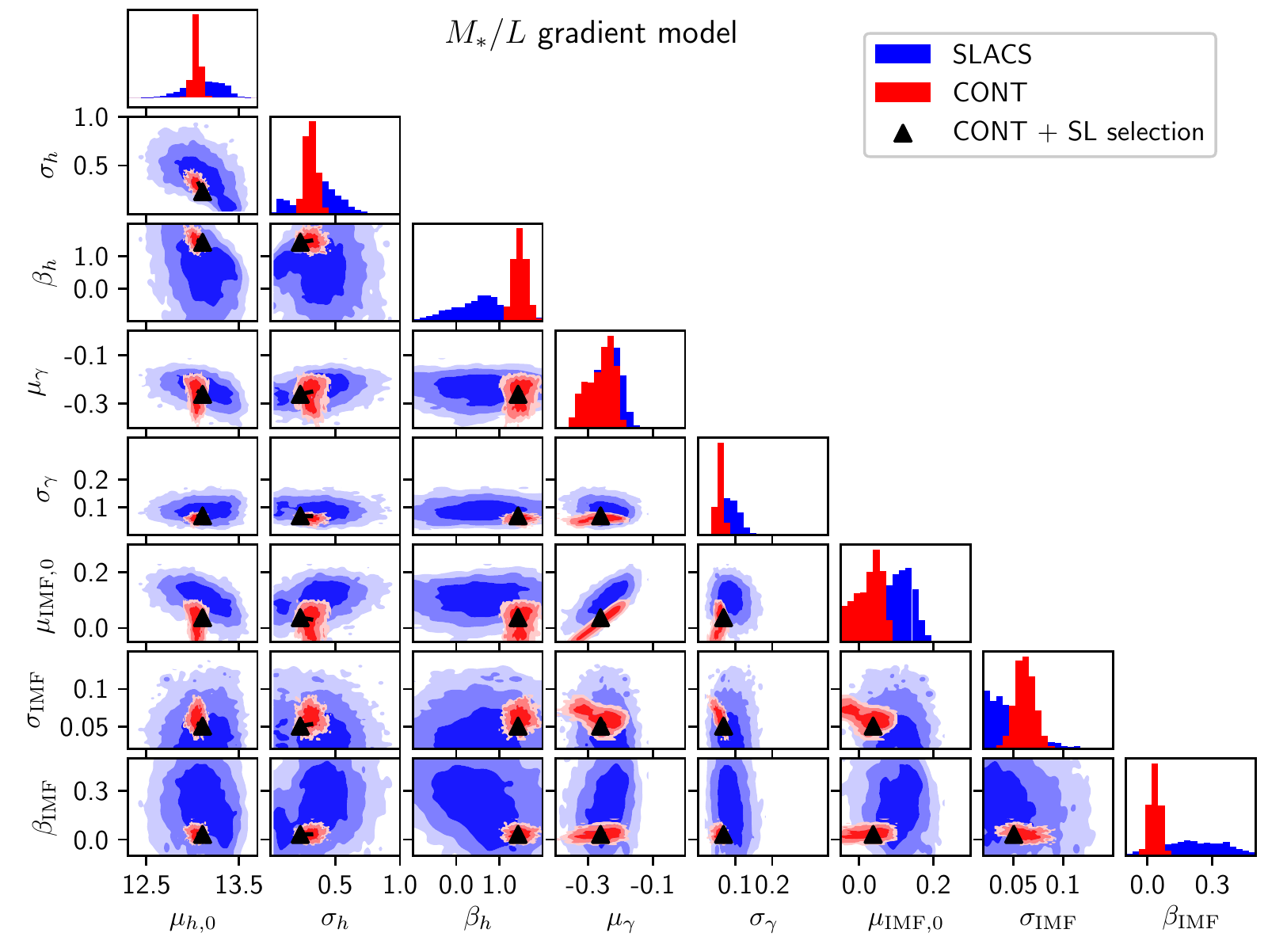}
\caption{$M_*/L$ gradient model. Posterior probability distribution of the parameters describing the distribution in halo mass, IMF mismatch parameter and IMF gradient. {\em Red region:} inference from the control sample. {\em Blue region:} inference from SLACS galaxies.
{\em Black triangles:} values of the hyper-parameters obtained by applying a strong lensing selection correction to the maximum-likelihood control sample model, as described in subsection \ref{ssec:lenscorr}.
 \label{fig:mlgradcp}
}
\end{figure*}

We fit the model to SLACS lenses and the control sample. The inference on the hyper-parameters describing halo mass, stellar IMF and $M_*/L$ gradient is plotted in \Fref{fig:mlgradcp}, while the median and 68\% enclosed probability values of all hyper-parameters are listed in \Tref{tab:mlgrad}.
This time there is good agreement between the inference based on SLACS data and on HSC weak lensing data, with or without strong lensing selection correction.
Both datasets find the need for a gradient in $M_*/L$, an average log halo mass of $\mu_{h,0}\approx 13$ and an average IMF intermediate between Chabrier and a Salpeter IMF (which would correspond to $\mu_{\mathrm{IMF},0} = 0.25$).
The inference is consistent with no mass dependence of the IMF (parameter $\beta_{\mathrm{IMF}} \sim 0$).
The agreement between the two inferences can also be seen by looking at the predicted average density profile, plotted in the bottom panel of \Fref{fig:rho_merged}.
\begin{table*}
\caption{$M_*/L$ gradient model. Median values and 68\% confidence interval of the posterior probability distribution of individual hyper-parameters, marginalised over the rest of the hyper-parameters.}
\label{tab:mlgrad}
\begin{tabular}{lcccl}
\hline
\hline
 & SLACS & Control & Control, SL pred. & Parameter description \\
\hline
$\mu_{h,0}$ & $13.12 \pm 0.19$ & $13.04 \pm 0.04$ & 13.11 & Average $\log{\mhalo}$ at stellar mass $\log{\mchab}=11.3$\\
$\sigma_h$ & $0.37 \pm 0.16$ & $0.31 \pm 0.04$ & 0.23 & Dispersion in $\log{\mhalo}$ around the average\\
$\beta_h$ & $0.54 \pm 0.69$ & $1.48 \pm 0.15$ & 1.43 & Power-law dependence of halo mass on $\mchab$\\
$\mu_\gamma$ & $-0.24 \pm 0.04$ & $-0.26 \pm 0.05$ & -0.26 & Average $M_*/L$ gradient parameter\\
$\sigma_\gamma$ & $0.09 \pm 0.02$ & $0.06 \pm 0.01$ & 0.07 & Dispersion in the gradient parameter\\
$\mu_{\mathrm{IMF}, 0}$ & $0.11 \pm 0.04$ & $0.02 \pm 0.04$ & 0.04 & Average $\log{\aimf}$ at stellar mass $\log{\mchab}=11.3$\\
$\sigma_{\mathrm{IMF}}$ & $0.05 \pm 0.02$ & $0.06 \pm 0.01$ & 0.05 & Dispersion in $\log{\aimf}$ around the average\\
$\beta_{\mathrm{IMF}}$ & $0.23 \pm 0.15$ & $0.03 \pm 0.02$ & 0.03 & Power-law dependence of IMF normalization on $\mchab$\\
$\mu_*$ & $11.28 \pm 0.09$ & $11.24 \pm 0.03$ & 11.34 & Average parameter in \Eref{eq:fullskew}\\
$\sigma_*$ & $0.22 \pm 0.05$ & $0.23 \pm 0.01$ & 0.24 & Dispersion parameter in \Eref{eq:fullskew}\\
$\log{s_*}$ & $-0.44 \pm 0.86$ & $-0.31 \pm 0.16$ & -0.30 & Log of the skewness parameter in \Eref{eq:skew}\\
$\rm{Median}\,\log{M_*}$ & $11.37 \pm 0.03$ & $11.33 \pm 0.01$ & 11.43 & Median stellar mass (not a hyper-parameter) \\

\hline
\end{tabular}
\end{table*}

The goodness of fit is also improved, compared to the previous two models, as can be seen by comparing the bottom rows of \Fref{fig:deltasigma_pp} and \Fref{fig:rein_sigma_gammap} with the middle and top panels. In particular, the range of values of the inner density slope $\gamma'$ predicted by the $M_*/L$ gradient model is now a good match to the observations.

By fitting the posterior predicted $\gamma'$ distribution with a Gaussian with mean dependent on stellar surface mass density, 
\begin{equation}
\pr(\gamma'|\Sigma_*) = \frac{1}{\sqrt{2\pi}\sigma_{\gamma'}}\exp{\left\{-\left(\frac{(\gamma' - \mu_{\gamma'}(\Sigma_*))^2}{2\sigma_{\gamma'}}\right)\right\}},
\end{equation}
\begin{equation}
\mu_{\gamma'} = \gamma_0 + \eta_{\gamma'}(\log{\Sigma_*} - 9.0),
\end{equation}
we find an average $\gamma'$ at the pivot surface mass density $\gamma_0 = 2.13$, a dependence of $\gamma'$ on $\Sigma_*$ of $\eta_{\gamma'} = 0.32$, and an intrinsic scatter $\sigma_{\gamma'} = 0.09$.
The values inferred by \citet{Son++13b} for the SLACS sample are $\gamma_0 = 2.11\pm0.02$ (taking into account the redshift-dependence of $\gamma'$), $\eta_{\gamma'} = 0.38\pm0.07$, and $\sigma_{\gamma'} = 0.12\pm0.02$, in good agreement with the model prediction.

The predicted distribution in Einstein radius is closer to the observed one, with respect to previous models, although still unable to reproduce the sharp peak around $1.2''$. This suggests that either the model or the strong lensing selection correction are not a perfect description of reality. Nevertheless, the $M_*/L$ gradient model performs significantly better compared to the vanilla and the adiabatic contraction, and from now on we will adopt it as our fiducial model.

There is a strong degeneracy between the parameter describing the average stellar IMF normalisation, $\mu_{\mathrm{IMF},0}$ and the one describing the average $M_*/L$ gradient, $\mu_\gamma$, for both the SLACS and the control sample datasets (see panel in row 6 and column 4 in \Fref{fig:mlgradcp}).
This is because, as we explained in subsection \ref{sect:vanilla}, the parameters describing the stellar component are mostly constrained by stellar velocity dispersion data, for which we only have single aperture measurements from SDSS. With only one stellar kinematics data point per galaxy, it is not possible to constrain both the IMF normalisation and the $M_*/L$ gradient on individual objects. This degeneracy can be broken only partially with the statistical combination of measurements over many galaxies.

\section{Discussion}\label{sect:discuss}

We fitted models for the structure of massive ETGs to two sets of massive quiescent galaxies, one consisting of strong lenses and the other one drawn from the general population of galaxies, using weak lensing, strong lensing and stellar kinematics data.
We considered three different models. The first two models, consisting of a de Vaucouleurs stellar bulge and a dark matter halo with varying degrees of freedom in the density profile, fail to self-consistently describe strong lenses and the sample of massive galaxies, even after accounting for lensing selection effects.
These results highlight the challenge of finding an accurate description for the density profile of massive galaxies over the two decades in radius probed by our data, $1\,\rm{kpc} < R < 300\,\rm{kpc}$.

One of the requirements for a successful model is the ability to reproduce the range of total density slopes measured for the strong lenses. About a third of the SLACS lenses have a total density profile steeper than what can be accounted for with mass-follows-light models ($\gamma' > 2.2$). 
The only dark matter models that would be able to explain these observations are the ones for which the inner density profile is allowed to be steeper than that of the stars. However, we reject such models based on our physical prior that there are no known mechanism that can produce such steep dark matter profiles.

A model for the mass distribution consisting of NFW halos and de Vaucouleurs profiles with a radial gradient in stellar mass-to-light ratio is instead able to describe SLACS strong lenses and the control sample of massive galaxies with the same sets of hyper-parameters.
The maximum-likelihood value of the average $M_*/L$ gradient slope found for SLACS galaxies is $\mu_\gamma=-0.24$. 
Is this value of the slope reasonable?
How much of this gradient can be attributed to gradients in stellar population properties at fixed IMF, and how much could be due to a gradient in IMF?

\begin{figure}
\includegraphics[width=\columnwidth]{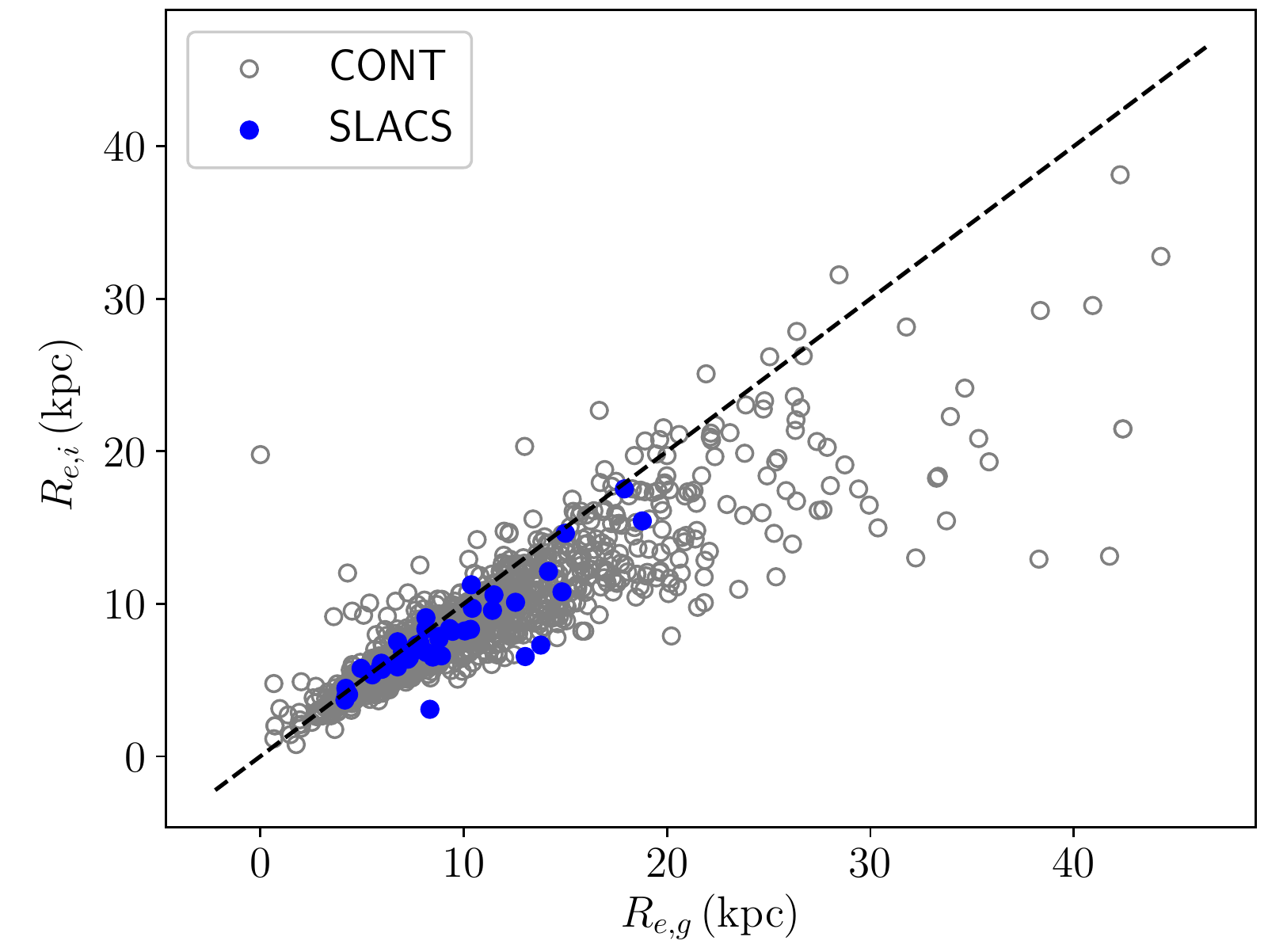}
\caption{Observed frame $i$-band effective radius as a function of $g$-band effective radius, from SDSS photometry, for SLACS lenses and the control sample.
}
\label{fig:colorgrad}
\end{figure}

The galaxies in both our samples show colour gradients: they are redder in the centre with respect to the outskirts. This can be seen by looking at the distribution of effective radii measured in different photometric bands, plotted in \Fref{fig:colorgrad}: half-light radii in redder bands are systematically smaller than in bluer bands.

Colour gradients can be the result of a gradient in age, dust content or metallicity.
Massive ETGs such as the objects in our samples have typically very little dust. This leaves age and metallicity as the most probable origins of the gradients.
Varying age at fixed metallicity or vice-versa results in a higher $M_*/L$ in correspondence with redder colour.
Therefore, regardless of the interpretation, the colour gradients observed in our sample are qualitatively in agreement with the negative $M_*/L$ slopes we inferred dynamically.

In the literature, there are more quantitative estimates of $M_*/L$ gradients in ETGs.
\citet{Tor++11} carried out a stellar population synthesis analysis on a sample of $\sim$10,000 galaxies with SDSS photometry, allowing for gradients in age and metallicity in their model.
They found that massive ETGs show on average positive $M_*/L$ gradients, i.e. a higher mass-to-light ratio in the outskirts with respect to the central parts, in contrast with our inference, although they find negative gradients if they limit the analysis to systems with older stellar populations.
 
\citet{Szo++13} used an empirical colour-$\Upsilon_*$ relation and SDSS photometric data to constrain gradients in mass-to-light ratio on a sample of 220 high mass galaxies with HST data.
They quote an average ratio between half-mass radius and half-light radius of $R_m/R_e\sim0.75$ for this sample. This value corresponds to a mass-to-light ratio slope of $\gamma\approx-0.13$, as can be seen from \Fref{fig:upsilon}.

\citet{New++15} estimated $M_*/L$ gradients in a sample of ten strong lens ETGs in the centre of massive groups, converting colour gradients into $M_*/L$ gradients assuming a fixed age. They found a median $M_*/L$ slope of $-0.15$.

Finally, \citet{PCM17} carried out a study of the radial profile of $M_*/L$ of galaxies in the Atlas 3D sample \citep{Cap++11}, using spatially resolved spectroscopic data.
They found an average $M_*/L$ profile very close to flat.

Overall, these studies reveal mass-to-light ratio gradients that are somewhat shallower than what is required to account for the average slope we measured from SLACS data, $\mu_\gamma = -0.24\pm0.04$, although our inference on the slope is strongly degenerate with the IMF mismatch parameter, as we already pointed out.
The measurements listed above have been obtained with models with a fixed IMF, but steeper $M_*/L$ slopes can be obtained by 
allowing for a gradient in IMF.
If we allow for the IMF to be not universal, then we should expect IMF gradients in massive ETGs, given our understanding of their mass assembly history. The cores of these objects are believed to form at high redshift, while the outskirts are accreted via mergers with smaller systems down to present times. Stars in different regions of massive ETGs are then believed to have formed in different environments, and therefore could have different IMFs.
 
Radial gradients in the IMF have been claimed by \citet{Mar++15}, \citet{LaB++16}, and \citet{van++17}, based on spatially resolved spectroscopic studies of nearby ETGs. On M87, a gradient in IMF has been detected both via dynamical modelling \citep{O+A18a}, and from a spectral analysis \citep{Sar++18}.
\citet{Col++18} measured an IMF gradient in a nearby lens, through a joint lensing and stellar kinematics analysis.
A similar result was found for a cluster brightest central galaxy by \citet{SLE17b}, although they allow for an alternative interpretation consisting of a spatially constant IMF and a very massive central black hole.
At the moment, we cannot claim nor rule out IMF gradients based on our analysis.
In order to determine whether IMF gradients are needed to reproduce our measurements, we need 
to carry out a spatially resolved stellar population synthesis analysis of the galaxies in our samples.
We leave this for a future study.

Regardless of its origin, allowing for a gradient in mass-to-light ratio changes dramatically the average IMF inferred from the combination of lensing and stellar kinematics.
While \citet{Aug++10b} and \citet{Son++15} found an average IMF normalisation heavier than a Salpeter IMF in their study of SLACS lenses, our current analysis yields a value of $\mu_{\mathrm{IMF},0}=0.02\pm0.04$, consistent with a Chabrier IMF. 
In parallel, the inferred dark matter masses are revised upwards.
\citet{Son++15} measured values of the projected dark matter mass fraction within $5\,\rm{kpc}$ below 20\%, significantly smaller than predictions from numerical simulations \citep[see e.g. Figure 10 of][]{Xu++17}.
In the current analysis, the dark matter fraction for the same objects is increased to 30\%.

\begin{figure} 
\includegraphics[width=\columnwidth]{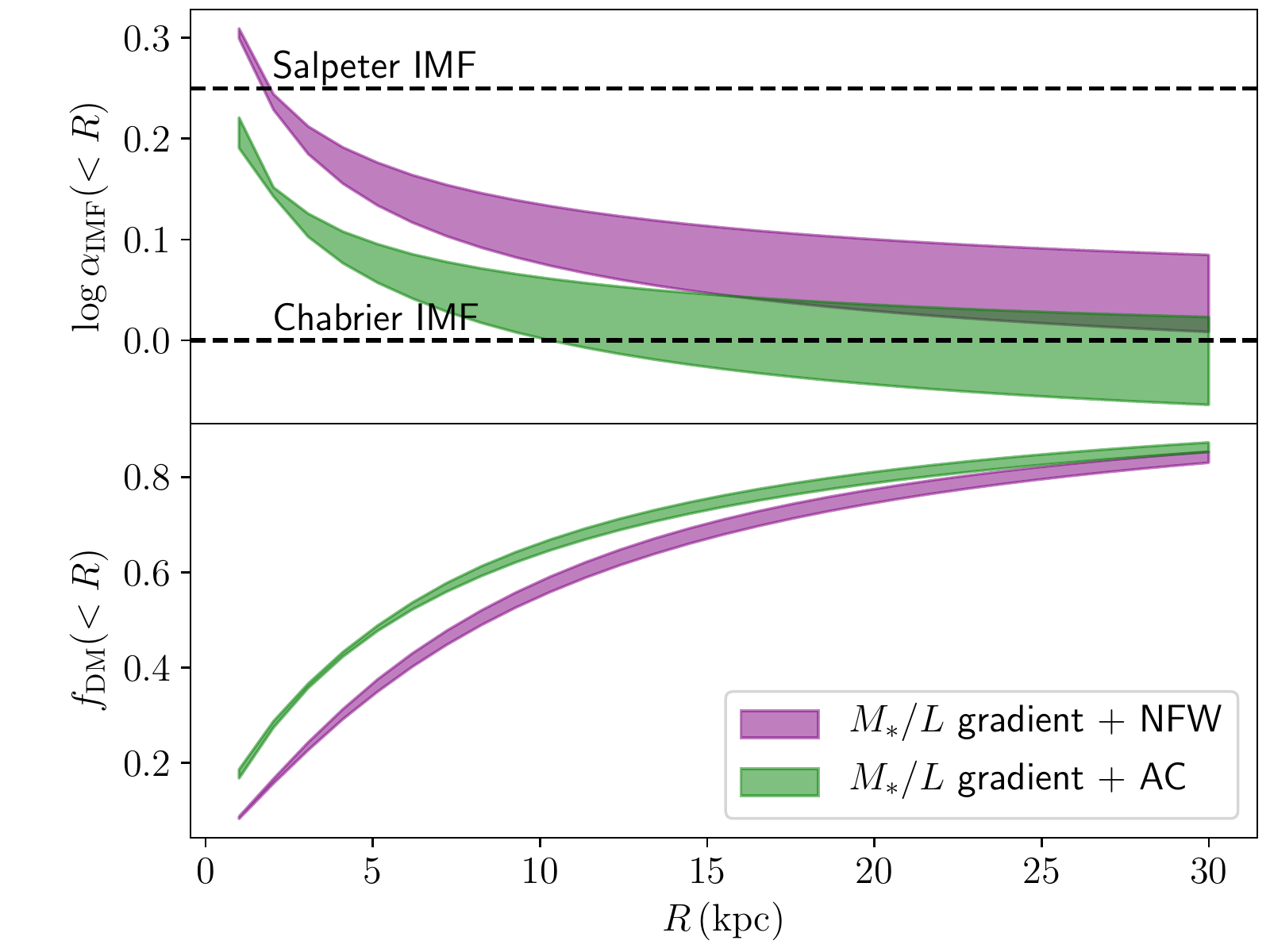}
 \caption{
{\em Top:} enclosed IMF normalization profile, defined as the ratio between the true stellar mass and the stellar population synthesys stellar mass, calculated as a function of projected distance from the center, for a galaxy with $\log{\mchab}=11.3$ and $\reff=7\,\rm{kpc}$. The band spans the $1\sigma$ confidence region, as inferred by fitting the control sample.
The purple band corresponds to our fiducial model, introduced in subsection \ref{sect:mlgrad}, while the green band is inferred by fitting a model with a $M_*/L$ gradient and dark matter halos described by an adiabatically contracted profile with contraction efficiency $\nu=0.5$.
{\em Bottom:} projected dark matter fraction profile for the two models.
}
\label{fig:alphaenc}
\end{figure}

We point out, however, that the value of the IMF normalization we infer is particularly sensitive to two of our model assumptions: the functional form of the $M_*/L$ profile, which we assume to be a powerlaw at all radii (\Eref{eq:grad}), and the choice of the dark matter density profile, which in our fiducial model is fixed to an NFW model.
To gauge the impact of the assumed $M*/L$ gradient, we plot in the top panel of \Fref{fig:alphaenc} the inferred enclosed IMF mismatch parameter profile, $\aimf(<R)$, defined as the ratio between the true stellar mass enclosed within projected radius $<R$ and the stellar population synthesys-derived stellar mass enclosed within the same aperture, for an average galaxy with $\log{\mchab}=11.3$ and $\reff = 7\,\rm{kpc}$. 
For our fiducial model (purple contour), the value of $\aimf(<R)$ converges to the inferred value of $\mu_{\mathrm{IMF},0}=0.02\pm0.04$ at large radii. However, around the scale of the half-light radius, $5\,\rm{kpc} < R < 10\,\rm{kpc}$, the region our dynamical constraints are most sensitive to, we infer values of $\aimf(<R)$ that are $\sim0.1$~dex higher, roughly in between the values corresponding to a Chabrier and a Salpeter IMF.
Since the stellar mass profile is relatively unconstrained at large radii, we consider the inferred value of $\aimf$ in the inner regions to be more robust, compared to the value of $\mu_{\mathrm{IMF},0}$: an alternative model in which the $\aimf(<R)$ profile remains flat at $R > 10\,\rm{kpc}$ would most likely be indistinguishable from our model based on a pure power-law $M_*/L$ profile, given our data.

To estimate how much our results depend on the assumed dark matter density profile, we repeat the analysis assuming adiabatically contracted dark matter halos, following the model introduced in subsection \ref{sect:adcontr}, with a value of the contraction efficiency parameter fixed to $\nu=0.5$.
We pick this value because it is close to the average contraction efficiency we inferred when fitting the control sample with the model with no $M_*/L$ gradient.
The main effect of assuming a dark matter model with a steeper inner density profile is a change in the value of the average IMF normalization, which for the control sample is inferred to be $\mu_{\mathrm{IMF},0}=-0.06\pm0.04$.
This change is illustrated in the top panel of \Fref{fig:alphaenc}, where the enclosed $\aimf$ profile for this new model is plotted in green.
We point out how, while the average IMF normalization goes down in a model with adiabatically contracted dark matter halos, the inferred value of the $M_*/L$ gradient does not change significantly. The need for a gradient in the model appears to be robust to the particular choice of the dark matter density profile.

In the bottom panel of \Fref{fig:alphaenc} we plot the inferred projected dark matter fraction profile for the two models.
This quantity is, quite naturally, particularly sensitive to the assumed dark matter density profile.

Relaxing the assumption of a spatially constant $M_*/L$ is the key ingredient for reconciling strong lensing observations, which probe the mass distribution at the scales of a few kpc, with weak lensing constraints, probing scales in the range $10\,\rm{kpc}$-$300\,\rm{kpc}$.
In principle there are other model assumptions that, if relaxed, could help mitigate the tension between the SLACS and HSC measurements.
As is often done in joint strong lensing and stellar dynamics studies, we assumed spherical symmetry for the gravitational potential and isotropic orbits for the stars.
If we allow, for example, for radial anisotropy, the same mass distribution would produce a higher line of sight velocity dispersion.
As pointed out earlier, failure to match the velocity dispersion of some SLACS lenses is the reason why fits with our vanilla and adiabatically contracted/expanded model return very low halo masses.
Therefore, in principle, allowing for radial anisotropy could help bring halo masses inferred from strong lensing closer to the values inferred from weak lensing.

Dynamical models that allow for anisotropy have been fitted to SLACS data by \citet{Pos++15}. Even by allowing for anisotropy, the inferred dark matter fractions are very close to zero for a significant number of objects, suggesting that anisotropy alone is unlikely to be the solution to the apparent tension between SLACS and HSC measurements observed in the context of the vanilla model.

Another effect that could increase the predicted velocity dispersions is allowing for elongation in the mass distribution along the line of sight \citep[see e.g. Figure 17 of][]{Son++12}.
Deviations from spherical symmetry of SLACS lenses have been explored by \citet{Bar++11} using spatially resolved stellar kinematics measurements.
They found that the inferred density profiles are consistent with those derived from a simple spherical Jeans equation approach.

The presence of a central supermassive black hole, not included in our fiducial model, would also boost the central velocity dispersion, increasing the mass enclosed in the central regions in a similar way as a radially decreasing IMF normalization does \citep[see e.g.][]{SLE17a}.

To test the effects of supermassive black holes on our measurement, we consider a `vanilla + black hole' model, consisting of NFW dark matter halos, de Vaucouleurs stellar profiles with spatially constant $M_*/L$ and a central black hole with mass $\mbh$. 
Fitting individual SLACS lenses with this model, assuming uninformative (flat) priors on the model parameters, can lead to very large, unphysical, values of the ratio between black hole mass and stellar mass, especially for the lenses with steep density slopes ($\gamma' > 2.3$).
We then put a prior on $\mbh$ assuming a power-law relation between black hole mass and the central velocity dispersion, with Gaussian scatter in $\log{\mbh}$ at fixed velocity dispersion, as measured by \citet{M+M13}.
We also set an upper limit on the ratio between black hole mass and stellar mass: $\mbh/\mstar < 0.01$.

We fit the SLACS sample with this model, using the same form for the distribution of halo mass, stellar mass and IMF normalization as that used to describe the vanilla model. We obtain an averagae log-halo mass of $\mu_{h,0} = 12.51\pm0.19$. This is slightly larger than the value obtained with the vanilla model, but still inconsistent with the halo mass constraints from weak lensing. We then conclude that accounting for the presence of central black holes does not change significantly our inference.

Finally, some of our results could, in principle, change if we drop the assumption of a de Vaucouleurs profile for the light distribution. For instance, adopting the more general S\'{e}rsic model \citep{Ser63} could lead to overall steeper density profiles, that would in part mimic the effects of a mass-to-light ratio gradient. 
\citet{Sha++18} showed how an evolving S\'{e}rsic index with time can help match simple semi-empirical models with lensing observations of the redshift evolution of the density slope $\gamma'$.
In order to robustly assess the impact of the use of a S\'{e}rsic model, we need to obtain self-consistent stellar mass, S\'{e}rsic indices and effective radii for both the SLACS lenses and the control sample. We leave this exploration for future work.
Nevertheless, the more flexible analysis of \citet{Pos++15}, who modelled the surface brightness distribution of SLACS lenses with a multi-Gaussian expansion, produced similar results to our study, suggesting that the effect of deviations from a pure de Vaucouleurs profile are probably small.

We point out how very low inferred dark matter fractions are not a peculiarity of our study of SLACS lenses, but appear to be a common occurrence in stellar dynamics measurements on ETGs, even when more sophisticated modelling tools compared to our isotropic spherical Jeans analysis are used. 
This is the case, for example, for the analysis of Atlas 3D galaxies by \citet{Cap++13}.
\citet{Cap++13} used Jeans axisymmetric dynamical models with orbital anisotropy to fit stellar kinematics data from integral field spectroscopy.
They found a median dark matter fraction within the 3D half-light radius as low as 13\%, 
indicating that low inferred dark matter masses are robust to the inclusion of anisotropy or deviations from spherical symmetry.
The \citet{Cap++13}, however, assumed a spatially constant mass-to-light ratio for the stellar component.
As pointed out by \citet{Ber++18}, allowing for a gradient would change significantly the inferred stellar and dark matter masses of the Atlas 3D sample of galaxies, as is the case for our analysis of SLACS lenses.
Recently, \citet{Pec++17} re-analysed a subsample of 27 low-mass ($\log{M_*} < 10.5$) galaxies from the Atlas 3D sample, using high resolution imaging from HST. They found evidence for a increase in the stellar mass-to-light ratio in the inner regions of these galaxies, which they interpret as the combined effect of the presence of over-massive central black holes and IMF gradients.

\subsection{Differences between strong lenses and non-lenses}

A fundamental feature of our analysis is the treatment of strong lensing selection, needed to compare the inference based on the SLACS sample of lenses with the control sample, and described in subsection \ref{ssec:lenscorr}.
Regardless of the model used, a common result of strong lensing selection is to increase the average stellar mass of the sample.
Lensing selection has instead a very limited effect on the distribution of halo masses at fixed stellar mass.
As discussed in subsection \ref{sect:vanilla}, the reason for this is the low sensitivity of the lensing cross-section to the halo mass. The lensing cross-section is determined by the projected mass within a few kpc, a region dominated by the baryons.
As a result, the main effect of strong lensing selection is to modify the distribution in stellar mass.

The change in IMF normalisation due to lensing selection is also quite small, in our fiducial model with mass-to-light ratio gradients. This is because the inferred value of the intrinsic scatter in the IMF normalisation is small, $\sigma_{\mathrm{IMF}} = 0.06\pm0.01$. With a small intrinsic scatter, the effect of lensing selection is also small.

\begin{figure} 
\includegraphics[width=\columnwidth]{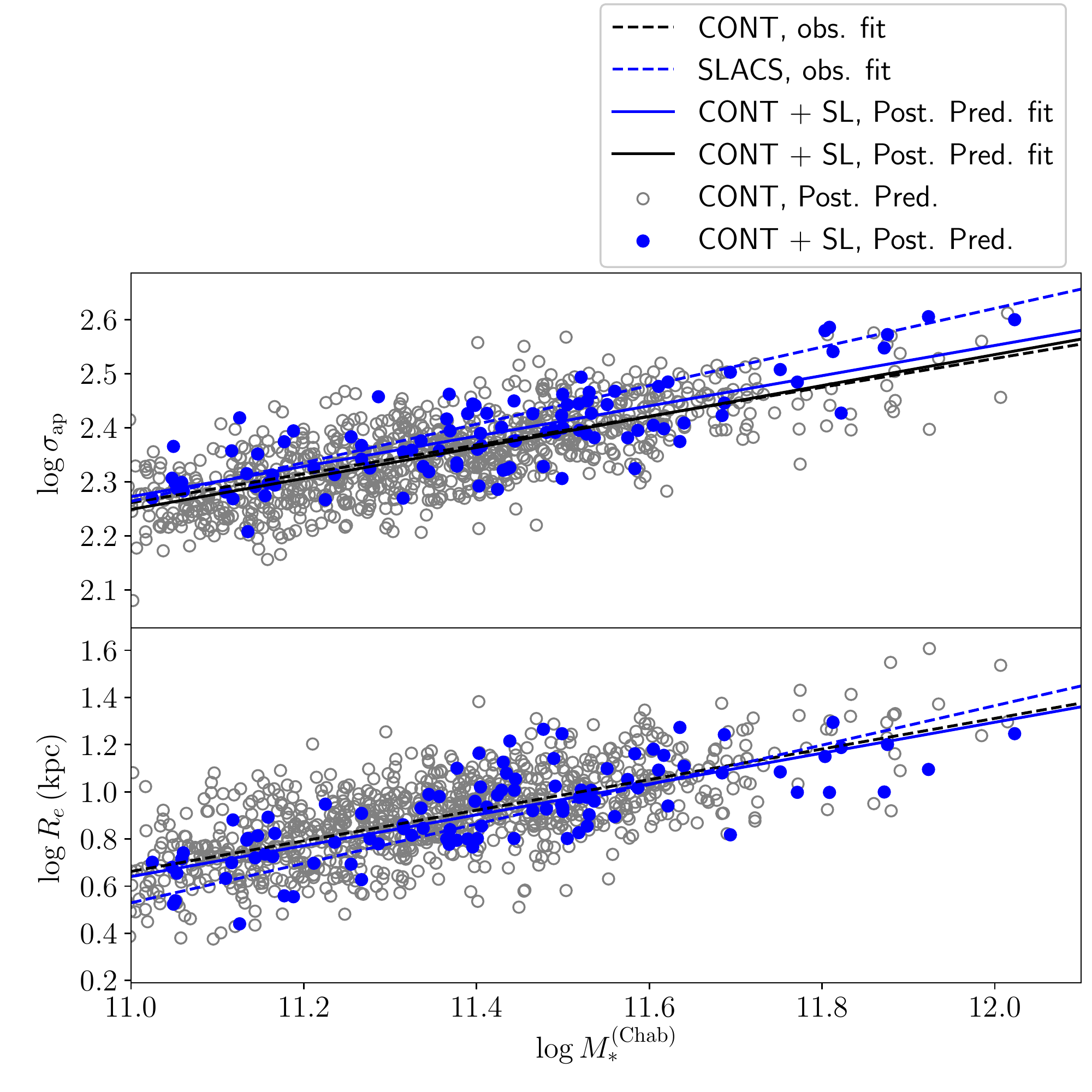}
 \caption{Stellar mass vs. effective radius and velocity dispersion of mock galaxies.
{\em Gray circles:} Sample of 1,000 mock galaxies generated from the maximum-likelihood point of the $M_*/L$ gradient model, fitted to the control sample.
{\em Blue circles:} Subsample of 100 objects selected with a random draw weighted by the strong lensing cross-section of each galaxy.
{\em Dashed lines:} observed best-fit $\mchab-\sigmaap$ (top) and $\mchab-\reff$ relation (bottom) for the SLACS and the control sample, as plotted in \Fref{fig:mstarreffsigma}.
{\em Solid lines:} best-fit $\mchab-\sigmaap$ (top) and $\mchab-\reff$ relation (bottom) inferred from the posterior predictive sample, with (blue) or without (black) the strong lensing selection correction.
The $\mchab-\reff$ relation of the mock sample is omitted from the bottom panel, since it is set equal to the observed relation by construction.
}
\label{fig:postreff}
\end{figure}

The mass-size and mass-velocity dispersion relation are also relatively unchanged by strong lensing selection.
This is shown in \Fref{fig:postreff}, where we plot the $\mchab-\reff$ and $\mchab-\sigma$ relation of the mock sample, generated from the fiducial model fitted to the control sample, together with the distribution of a sample of lenses, drawn from the same mock sample with a probability proportional to the strong lensing cross-section of each galaxy.
Strong lenses are only slightly more compact, $\sim0.02$~dex on average, at fixed stellar mass, with respect to the general population, and have higher velocity dispersion. 
This is in qualitative agreement with observations of the SLACS sample, although we are unable to reproduce in detail the observed relations, particularly the smaller average sizes of SLACS lenses observed at the low-mass end.
However, the difference between our prediction and the observed $\mchab-\reff$ relation is of the same magnitude as the 8\% systematic uncertainty on our measurements of $\reff$, and therefore not very significant.

%

\subsection{Model prediction: velocity dispersion profile}

The $M_*/L$ gradient model provides a good fit to the data used in this study: Einstein radii, central velocity dispersions and, on larger scales, weak lensing.
In principle, we could add more constraints to the analysis, in order to test even more complex models than the current one.
One possibility would be to make use of spatially resolved kinematics measurements, which could, for example, allow us to relax the assumption of isotropic orbits.
Although there are integral field spectroscopy observations available for 17 of the SLACS lenses \citep{Czo++12}, their analysis would require a significant amount of work, and we leave it for future study.
Nevertheless, we can use our model to predict the velocity dispersion profile of a typical galaxy, and make a qualitative comparison with existing measurements.

\begin{figure}
\includegraphics[width=\columnwidth]{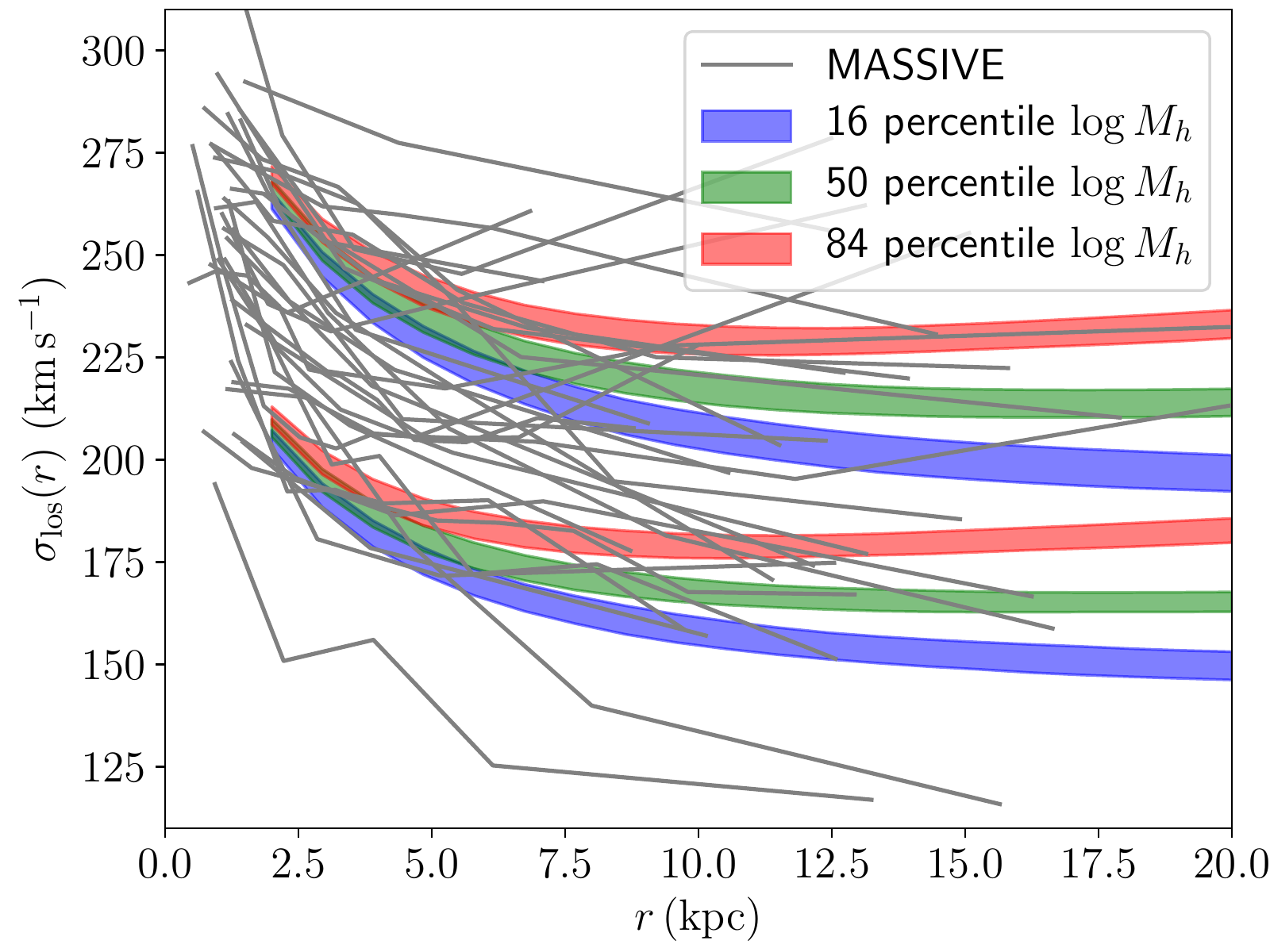}
\caption{{\em Shaded regions:} Predicted average line-of-sight velocity dispersion profile, calculated for two galaxy with $\log{\mchab}=11.3,\,11.6$, $\reff=7,\,10\,\rm{kpc}$, with average values of the IMF mismatch parameter and $M_*/L$ gradient slope, and halo mass corresponding to the 16, 50 and 84 percentile of the distribution, as inferred from the control sample.
The bands delimit 68\% enclosed probability regions.
The profiles are calculated using the spherical Jeans equation under the assumption of isotropic orbits.
{\em Grey lines:} velocity dispersion profiles measured by \citet{Vea++18} for a set of early-type galaxies from the MASSIVE survey, with $M_K > -25.6$.
\label{fig:idefix}
}
\end{figure}

In \Fref{fig:idefix}, we plot the line-of-sight velocity dispersion profile of two galaxies with $\log{\mchab}=11.3,\,11.6$, $\reff=7,\,10\,\rm{kpc}$, average IMF normalisation and $M_*/L$ gradient slope for their stellar mass, and halo mass corresponding to the 16, 50 and 84 percentile of the inferred distribution. We assume orbital isotropy. 
The predicted profile shows a relatively sharp drop in the line-of-sight velocity dispersion, followed by either a slowly falling or rising outer profile, depending on the halo mass.
The integral field spectroscopy data of SLACS lenses show a qualitatively similar behaviour in the inner regions \citep[see Figure 5 of][]{Czo++12}, but lack observations at radii larger than $\reff$, where the profile is predicted to flatten.

We can then compare our prediction with observations of $z\sim0$ early-type galaxies.
\citet{Vea++18} studied in detail the velocity dispersion profile of 90 early-type galaxies from the MASSIVE survey \citep{Ma++14}. Although the average mass of the \citet{Vea++18} sample is somewhat larger than that of our samples, there is significant overlap.
We over-plot in \Fref{fig:idefix} the velocity dispersion profile of a subset of MASSIVE galaxies, selected by setting a limit on their $K-band$ magnitude, $M_K > -25.6$. These are the objects plotted in the left-hand panel of Figure 1 of \citet{Vea++18}.
Most of the MASSIVE galaxies exhibit a velocity dispersion profile that is falling with radius in the inner regions and then rising or staying constant at larger radii, in qualitative agreement with our prediction. 

However, the \citet{Vea++18} reveals a much wider variety of velocity dispersion profiles compared to what our model is able to describe: there are objects with flat inner profile, and objects with steeply rising or falling outer profile, which are difficult to reproduce in the context of our model.
This seems to indicate that the real structure of massive galaxies is most likely more complex than our description.
The recent strong lensing study of \citet{O+A18b} appears to corroborate this idea. \citet{O+A18b} were able to disentangle the luminous and dark matter on individual objects for a set of 12 lenses. Even by allowing for the presence of $M_*/L$ gradients, which they detect in a quarter of their objects, they find large variations in the inner dark matter density slope across the population, ranging from cored density profiles to profiles as steep as isothermal.

It is possible that by relaxing our assumption of a universal dark matter density profile we can produce a broader range of velocity dispersion profiles, in better agreement with the observations from the MASSIVE survey. 
For a quantitative comparison between our model and spatially resolved kinematics data, however, additional ingredients, such as orbital anisotropy and rotation, must be taken into account.
%

\section{Conclusions}\label{sect:concl}

We analysed two samples of massive quiescent galaxies, selected in stellar mass by applying a cut $\log{\mchab} > 11.0$.
The first sample consists of $\sim1,700$ galaxies in the SDSS legacy spectroscopic sample, with weak lensing measurements from HSC.
The second sample consists of 45 strong lenses from the SLACS survey.
We fitted models for the distribution of stellar and dark matter mass and stellar IMF across the population of galaxies to weak lensing, strong lensing and stellar kinematics data on the two samples. 
We then compared the results obtained on the two samples, after correcting for strong lensing selection effects, with the goal of finding a simple model for structure of massive galaxies that can reproduce observations on both datasets.
We found the following.

\begin{itemize}
\item A model in which dark matter halos have an NFW density profile and the stellar mass follows the light distribution does not provide a good description of SLACS lenses: the inferred halo masses are too low compared to galaxies of the same stellar mass from the control sample, a discrepancy that cannot be ascribed to strong lensing selection.
\item Modifying the dark matter profile by allowing for adiabatic contraction or expansion does not solve the discrepancy between the two samples: SLACS lenses want strong adiabatic expansion while HSC galaxies prefer mild contraction.
This result is driven by lenses with a particularly steep total density profile, for which even mass-follows-light models struggle to fit both the strong lensing and stellar kinematics constraints.
\item Allowing for a gradient in stellar mass-to-light ratio, results in a good match between the SLACS sample and HSC galaxies, and shifts downwards the inferred average IMF normalization, the exact value depending on the assumed dark matter density profile.
\item In light of the current analysis, such a gradient can be interpreted both in terms of a gradient in stellar population properties at fixed IMF, or in terms of a gradient in IMF itself.
\item The main effect of strong lensing selection is to shift the median stellar mass towards higher values. Galaxy properties at fixed stellar mass show little differences between strong lens samples and the general galaxy population.
\end{itemize}

The main limitation of our study is the relatively low number of constraints for individual galaxies, which forces us to adopt rather simple mass models.
Another limitation is the necessity to correct for strong lensing selection effects when mapping the HSC sample, used for the bulk of our weak lensing analysis, to the SLACS sample of strong lenses.
The need for the use of two different samples of galaxies is dictated by the low signal-to-noise ratio of the weak lensing data available for the strong lens sample.
This situation can be avoided by using a much larger sample of strong lenses, so that both the strong and the weak lensing analysis can be carried out on the same set of objects. Current surveys such as HSC, the Kilo Degree Survey \citep{deJ++15} and the Dark Energy Survey \citep{Die++14}, are enabling the discovery of hundreds of new lenses \citep{Son++17, Pet++17, Die++17}, and in the future it will be possible to perform a joint strong and weak lensing analysis on a large sample of galaxies.

\section*{acknowledgments}
We thank Melanie Veale for kindly providing observed velocity dispersion profiles of MASSIVE galaxies.
This work was supported by World Premier International Research Center Initiative (WPI Initiative), MEXT, Japan.
AS is partly supported by KAKENHI Grant Number JP17K14250. 

\bibliographystyle{mnras}
\bibliography{references}

\end{document}